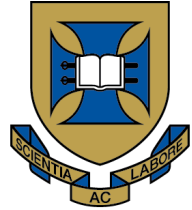

# THE UNIVERSITY OF QUEENSLAND
## AUSTRALIA

# Autonomous Penetration Testing using Reinforcement Learning

by

**Jonathon Schwartz**

School of Information Technology and Electrical Engineering,

University of Queensland.

Submitted for the degree of Bachelor of Science (Honours)

in the division of Computer Science.

Date of Submission

16th November, 2018

16th November, 2018

Prof Michael Brünig

Head of School

School of Information Technology and Electrical Engineering

The University of Queensland

St Lucia QLD 4072

Dear Professor Brünig,

In accordance with the requirements of the Degree of Bachelor of Science (Honours) majoring in Computer Science in the School of Information Technology and Electrical Engineering, I submit the following thesis entitled

"Autonomous Penetration Testing using Reinforcement Learning"

The thesis was performed under the supervision of Dr Hanna Kurniawati. I declare that the work submitted in the thesis is my own, except as acknowledged in the text and footnotes, and that it has not previously been submitted for a degree at the University of Queensland or any other institution.

Yours sincerely,

\_\_\_\_\_\_\_\_\_\_\_\_\_\_\_\_\_\_

Jonathon Schwarz

# Acknowledgements

I would like to acknowledge my supervisor, Dr Hanna Kurniawati, for her guidance and patience throughout this project as well as for the support she gave for the direction I took in the project. I would also like to acknowledge post-doctoral fellow in the robotics group, Troy McMahon, for his advice while working on this project. I would also like to thank the School of ITEE administrative staff for their help and coordination, which made this year much easier.

# Abstract

Penetration testing involves performing a controlled attack on a computer system in order to assess it's security. It is currently one of the key methods employed by organizations for strengthening their defences against cyber threats. However, network penetration testing requires a significant amount of training and time to perform well and presently there is a growing shortage of skilled cyber security professionals. One avenue for trying to solve this problem is to apply Artificial Intelligence (AI) techniques to the cyber security domain in order to automate the penetration testing process. Current approaches to automated penetration testing have relied on methods which require a model of the exploit outcomes, however the cyber security landscape is rapidly changing as new software and attack vectors are developed which makes producing and maintaining up-to-date models a challenge. To try and address the need for exploit models this project investigated the application of Reinforcement Learning (RL) to automated penetration testing. RL is an AI optimization technique that has the key advantage that it does not require a model of the environment in order to produce an attack policy, and instead learns the best policy through interaction with the environment. In the first stage of this study we designed and built a fast, light-weight and open-source network attack simulator that can be used to train and test autonomous agents for penetration testing. We did this by framing penetration testing as a Markov Decision Process (MDP) with the known configuration of the network as states, the available scans and exploits as actions, the reward determined by the value of machines on the network and using non-deterministic actions to model the outcomes of scans and exploits against machines. In the second stage of the project we used the network attack simulator to investigate the application of RL to penetration testing. We tested the standard Q-learning RL algorithm using both tabular and neural network based implementations. We found that within the simulated environment both tabular and neural network RL algorithms were able to find optimal attack paths for a range of different network topologies and sizes given only knowledge of the network topology and the set of available scans and exploits. This finding provided some justification for the use of RL for penetration testing. However, the implemented algorithms were only practical for smaller networks and numbers of actions and would not be able to scale to truly large networks, so there is still much room for improvement. This study was the first that the authors are aware of that has applied reinforcement learning to automated penetration testing. The versatility of RL to solving MDPs where the model is unknown, lends itself well to the changing nature of cyber security and could offer a valuable tool for reducing the workload on a cyber security professionals. Further work into developing scalable RL algorithms and testing these algorithms in higher fidelity simulators will be the next steps required before RL is ready to be applied in commercial environments.

# Contents

# List of Figures

# List of Tables

# Chapter 1

# Introduction

The increasing interconnection of the digital world via the use of the internet has changed the way businesses, governments and individuals operate and has lead to significant economic and social benefits [1]. This improved communication and availability, however, has also created more opportunities for cyber criminals to launch malicious attacks in the hopes of gaining access to sensitive data for their own gain [1]. These attacks can range in scale from massive state-sponsored attacks, such as the attempts to disrupt the US election, to simple attacks on individuals in the hopes of gaining password or credit card details for monetary gain [2], [3]. As more organisations and individuals come to rely on globally connected computer systems the ability to secure these systems against malicious attacks is becoming ever more important.

Cyber Security, which is the safeguarding of computer systems against unauthorized access or attack, is now a matter of global importance and interest [4]. Many nations have official policies in place regarding cyber security and some are investing significant amounts into the domain [4]. The Australian government only recently announced it will invest $50 million into cyber security research over the next seven years [5]. This increased focus and investment by governments and large organization emphasizes the serious threat cyber crimes pose for businesses, governments and individuals. It is important that effective methods and technologies are developed for securing computer systems against these threats.

One of the most commonly applied methods for evaluating the security of a computer system is penetration testing (pentesting). Pentesting involves performing an authorized controlled attack on a system in order to find any security vulnerabilities that could be exploited by an attacker [6], [7]. This method can be very effective for evaluating a systems security since it is essentially a simulation of what real world attackers would do in practice. This effectiveness, however, comes with one main drawback which is that it has a high cost in terms of time and skills required to perform it. This high cost is becoming more of an issue as digital systems grown in size, complexity and quantity which is causing a demand for security professionals, a demand that is not being met fast enough. In 2015 Cisco, one of the world's leading IT and networking companies, estimated there were more than 1 million unfilled security jobs worldwide [8]. Given this shortage of professionals and the necessity for pentesting in securing systems it is becoming crucial that tools and methods are developed for making pentesting more efficient.

Presently, a number of tools have been developed that facilitate penetration testers and help improve their efficiency. These tools include network and vulnerability scanners as well as libraries of known security vulnerabilities [8]. One of the most popular tools today is the open source Metasploit framework which has been in development since 2003 [8]. The Metasploit framework contains a rich library of known exploits of system vulnerabilities along with other useful tools such as scanners, which are used for information gathering on a target. Tools such as these allow the pentester to work at a higher level of abstraction where they are mainly focussed on finding vulnerabilities and selecting exploits rather than having to work at the low level of manually developing exploits. This enables pentesters to work faster and also makes security assessment more accessible to non-experts [9]. These tools have certainly been a boon to the cyber security industry, however, even with the great benefits these tools have provided they still rely on trained user, which are in short supply. Additionally, as systems grow in complexity the task of manually assessing security will become much harder to do systematically.

One approach to trying to solve the problem of conducting efficient and reliable pentesting is to apply techniques from the Artificial Intelligence (AI) planning domain to pentesting in order to automate the process. The original concept for this took the form of "attack graphs", which modeled an existing computer network as a graph of connected computers, where attacks can then be simulated on the network using known vulnerabilities and exploits [10]. Attack graphs can be effective in learning the possible ways an attacker can breach a system, however using these graphs requires complete knowledge of the system, which is unrealistic from a real world attackers point of view, and also require the manual construction of the attack graph for each system being assessed. Another approach taken involved modelling an attack on a computer as a Partially Observable Markov Decision Process (POMDP) [11], [12]. Modelling an attack as a POMDP introduces the attackers incomplete knowledge into the simulation and allows simulations to remove the assumption that the configuration of each host is known and instead models the observation of the configurations as the attack progresses. This approach can work well in practice against a single host machine but due to the computational properties of POMDPs, does not scale well [12]. In order to produce a method for automating pentesting that can handle the uncertainty inherent in any system while still being computationally feasible more research into novel methods is required.

A proposed solution to this problem is to simulate pentesting using an MDP. This approach would ignore the uncertainty about the state of the host computer's configuration and instead introduce the uncertainty into the success probability of each possible attack [13]. This type of model is computationally more feasible than the POMDP approach and does not require complete knowledge of the network and host configurations. However, it requires prior knowledge about the success probabilities of each possible action and it treats each target computer as exactly the same instead of utilizing information gathered about the target to produce more tailored attacks. Consequently, this approach addresses the issues of computational complexity and incomplete knowledge of network and host configurations but at the cost of accuracy of picking the best actions for each host.

Another approach for solving problems that can be framed as a MDP that does not require information about the transition model of the environment is Reinforcement Learning (RL) [14]. RL requires only the state space representation, the set of actions that can be performed and a reward function which defines what the RL agent is trying to achieve. The agent then learns a policy of actions to take from any given state through interaction with its environment. The use of RL has gained a lot of attention in recent years with its use in producing World Go champion beating agents [15], and although not as widely applied as other supervised machine learning approaches it has been successfully applied in the real world for in a number of robotics tasks [16]. RL provides a more general approach when a detailed model of the environment is not known.

In terms of automating penetration testing, we have information on the set actions that can be performed, and can frame the task into a reward function. However, due to the complex and ever changing nature of the cyber network environment, with constant updates to software and available exploits it becomes very difficult to maintain an accurate up-to-date model for the outcomes of performing any action. This combination sets RL as a good candidate approach for automated pentesting. However, RL offers its own challenges. It's generality comes at the cost of requiring a large amount of data in order to learn the best policy of actions. This data is typically gained from using simulations in order to train the RL agent.

In this study we aimed to investigate the use of RL in automated pentesting. The first part of the project was to develop a network attack simulator that can be used for testing and training of automated pentesting agents and algorithms. The second part of the study was to investigate the use of RL algorithms to finding policies for penetration testing in a simulated environment.

In this thesis, in chapter 2 we provide details on the current state of automated penetration testing and some background information useful for the later chapters of the thesis. Chapter 3 covers the rationale, design and implementation of the network attack simulator. In chapter 4 we provide a report on my investigation of using RL for automated penetration testing. Finally, chapter 5 provides a brief summary and conclusion of the the study and some directions for further research.

# Chapter 2

# Literature review

## 2.1 Penetration testing

Pentesting has been around for more than four decades and is a critical process in the development of secure cyber systems [17]. It involves performing a controlled attack on a piece of software or a network in order to evaluate its security. The process of pentesting, when applied to a network, is typically divided into a sequence of steps in order to methodically assess the given system (fig. 2.1.1). The specific steps can vary from attack to attack but generally a penetration test will first involve *information gathering*, where the aim is to find a vulnerability (a flaw in the system) in the network, followed by *attack and penetration* where the discovered vulnerability is exploited and then, lastly, using the newly gained access to repeat this process until the desired target is reached [9]. The information gathering step typically involves utilising tools such as traffic monitoring, port scanning and operating system (OS) detection in order to collect relevant information that can be used to determine if the system contains a vulnerability that can be exploited. The attack and penetrate phase involves executing a known *exploit*, which can be a program or certain data, to take advantage of the discovered vulnerability and to cause unintended behaviour in the system with the ultimate aim of compromising the target and gaining privileged access to it. Once a successful attack is performed the specific sequence of attack actions can then be reported and used by system administrators and developers to fix the vulnerabilities. Even though the systems and networks that are evaluated using pentesting can differ immensely, in each case the same general steps are followed. This has allowed for the development of a number of tools and frameworks to help make pentesting more efficient.

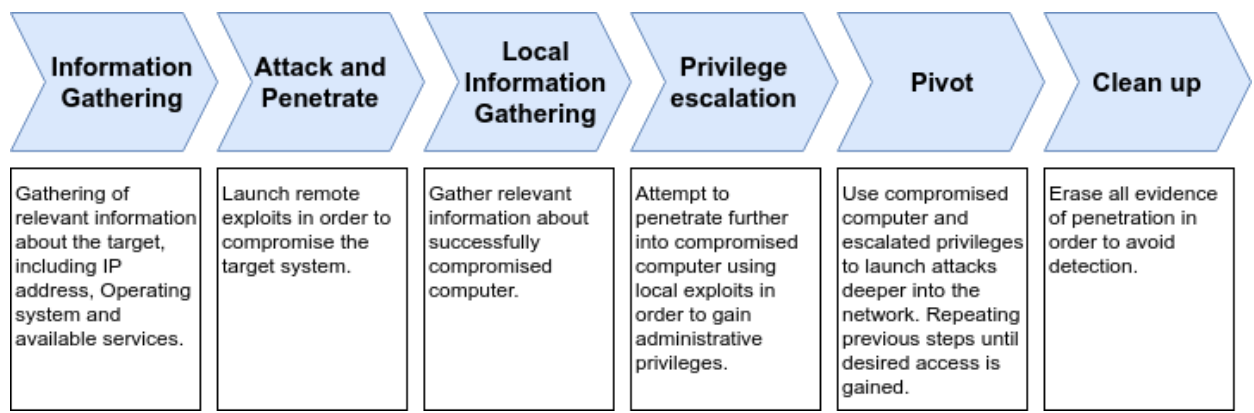

**Figure 2.1.1 | The main steps of a penetration test** [9].

## 2.2 Tools used for penetration testing

The cyber security industry is a very large and active community where there are numerous tools that exist to aid in pentesting, here we will mainly focus on the most commonly used tools for network pentesting. For the information gathering stage the aim is to find useful information about the network, this is typically done using a network scanner, with the best know network scanner being Nmap (www.nmap.org) [18], [19]. Nmap provides the user with information such as OS, open ports and services currently running on the system. This information can then be used to check for any vulnerabilities in the system using vulnerability scanners such as Nessus (www.tenable.com/nessus/professional) and, the free alternative, OpenVAS (www.openvas.org). These scanners can be used to determine if a vulnerability exists but in order to test the vulnerability an exploit needs to be used.

     Once information about vulnerabilities on system have been gathered, this is where pentesting frameworks such as Metasploit (www.metasploit.com) are used. The Metasploit software is a collection of tools and exploits that can be used from a single environment [18]. Metasploit is an open source project that was started in 2003 and was purchased by Rapid7 in 2009 and is regularly updated with new exploits and tools. The user can use information gathered during the information gathering stage to search for and launch exploits from the Metasploit software, allowing the user to focus on the tactical operation (which exploit to launch) rather than technical level (how to find and develop an exploit). This automation at the technical level allows for a huge increase in efficiency during pentesting.

     The tools now available to pentesters have had great benefits in terms of improved efficiency, however, it still requires a non-trivial amount of expertise and time to execute a

successful pentest. With the growing demand for security experts new approaches are necessary to ensure security evaluation demands can be met.

## 2.2 Automated pentesting: attack graphs

The idea of pentesting automation has been around for many years with the original approaches taking the form of *attack graphs* [10]. Attack graphs are used to model systems and how they are affected by specific exploits. In an attack graph, the nodes are typically the state of a system, where the state is defined by the current system configuration (i.e. OS, permissions, network connections, etc) and the edges connecting the nodes are known exploits [20]. Once this graph is constructed it becomes possible to search for sequences of attacker actions (exploits), known as an an *attack paths*, that lead to the attacker gaining unintended access to the system. Finding these attack paths can be done using classical AI planning techniques [21]. The main issue with this approach is that it requires complete knowledge of the network topology and each machines configuration, so is not realistic from an attackers point of view, and also requires manually setting up the graph for each new system being evaluated.

## 2.3 Automated pentesting: using MDP

Another approach to modelling and planning attacks against a system is to use a Markov Decision Process (MDP) to simulate the environment [13], [22]. A MDP is a general framework for modeling discrete decision making problems under uncertainty [23]. When defining an MDP we do so over the tuple $\{\mathcal{S}, \mathcal{A}, \mathcal{R}, \mathcal{T}\}$, where $\mathcal{S}$ is the state space, $\mathcal{A}$ is the action space, $\mathcal{T}$ is the transition function $\mathcal{T}(s, a, s') = P(s'|s,a)$ and $\mathcal{R}$ is the reward function $\mathcal{R}(s, a)$. At each time step the system will be in some state, $s \in \mathcal{S}$, and the agent will perform some action, $a \in \mathcal{A}$, resulting in two things: (1) a transition to a new state $s'$, where the new state is determined by the transition function $\mathcal{T}$, (2) a reward as determined by the reward function, $\mathcal{R}$. The aim of an agent attempting to solve a MDP is to find the optimal mapping from $\mathcal{S}$ to $\mathcal{A}$ so as to maximize the total accumulated reward. This mapping is known as the decision policy $\pi$.

When applied to pentesting the state space of the MDP becomes the possible configurations of the target machines or of the network, the actions are the possible exploits or scans available and the reward will be based on the cost of an action and the value gained when successfully compromising a system.

So far there has been limited application of MDPs to automated pentesting. One approach that has been used is to ignore the configuration of the target system entirely and instead rely on formulating the attackers uncertainty in the form of possible action outcomes [13]. Attacks are then planned based on the attack success probabilities, where each action receives a probability of success based on prior knowledge of the given action [13]. In this form it is essentially adding some non-determinism to an attack graph representation of a system.

The main advantage of this approach is that it allows for modelling of attacker uncertainty while still being computationally feasible to solve. However, this approach does not take into account known knowledge about the configuration of a system, which is a key step in effective penetration testing, and instead treats all machines as identical. Additionally, it also requires some prior knowledge of the attack outcome probabilities (the transition model) before it can be used and these probabilities can vary widely depending on the systems it is being used against (e.g. Windows or Linux) and will change over time as new software and exploits are developed.

## 2.4 Automated pentesting: modelling uncertainty with POMDP

Another approach to automating pentesting aimed to address the assumption of full network knowledge required for attack graphs while still accounting for the uncertainty of the attacker by is to model the pentesting problem as a POMDP [12].

A POMDP, or partially observable markov decision process, is a MDP in which there is uncertainty about the exact state the system is in and so it models the current state as a probability distribution over all possible states [24]. POMDPs are typically defined by a tuple $\{\mathcal{S}, \mathcal{A}, \mathcal{Q}, \mathcal{T}, \mathcal{O}, \mathcal{R}, b_o\}$ where $\mathcal{S}, \mathcal{A}, \mathcal{T}$ and $\mathcal{R}$ are the same as in an MDP, while $\mathcal{Q}$ is the observation space, $\mathcal{O}$ is the observation function $\mathcal{O}(s', a, o) = P(o|s',a)$ and $b_o$ is the initial probability distribution over states [24], [25]. Similar to an MDP, following each time step there will be a transition to a new state s' and a reward, but in addition taking a step will also result in an observation, $o \in \mathcal{Q}$, as determined by the observation function $\mathcal{O}$. The aim of the problem is the same as for an MDP, which is to find an optimal decision policy $\pi^*$.

When applied to pentesting the state space of the POMDP becomes the possible configurations of the target machine or of the network, the actions are the possible exploits or scans available, the observation space is the possible information that is received when a exploit or scan is used (e.g. ports open, or exploit failure/success), while the reward will be based on the cost of an action and the value gained when successfully compromising a system [11], [12]. Using the POMDP approach allows modelling of many relevant properties of real world hacking, since in a real attack the attacker would have incomplete knowledge about what the actual configuration of the system is and also whether certain exploits or scans will succeed. Additionally, it also means that the same automated pentesting approach can be used to test a system even if the system changes since knowledge of the exact system configuration is not assumed [11]. This differs when compared to attack graphs which would need to be updated every time something in the system is changed.

Sarraute et al. [12] implemented this approach using a POMDP in a simplified simulated network environment where the aim of the attack was to penetrate the system in order to reach a number of sensitive machines on the network. The approach was able to learn to intelligently mix scans and exploits and was tested on networks of varying size. Whether this approach performed better than the classical planning approach using attack graphs was uncertain but it definitely had the advantage of no assumed knowledge of the system configurations.

The POMDP approach is promising in that it more accurately models the real world attacker, however, it has one critical issue in that POMDP based solvers do not scale very well and quickly become computationally infeasible as the size of the state space grows. In the approach by Sarraute et al. [11], [12] they had to approach the network penetration test by decomposing the network into individual POMDPs against single target machines only. So although the approach was more realistic from attackers point of view it is currently computationally too expensive.

## 2.5 Automated pentesting: when the world model is unknown

Both MDP and POMDP provide a general framework for modeling the uncertainty inherent when performing penetration testing and have a number of techniques available for solving them. An MDP is the simpler version of the POMDP where the key difference is that there is no uncertainty about the current state of the environment (it is fully observable). The advantage of MDPs is that they are much more computationally tractable to solve when compared to POMDPs and there are many efficient algorithms for solving them [26]. This improved efficiency should allow them to be more useful in practice. However, this efficiency does come at the cost of some of the ability to model uncertainty. There main form of uncertainty that remains for MDPs is the uncertainty relating to non-deterministic actions as dictated by the transition function, or model of the world.

One technique that can be used for finding an optimal policy for a MDP is Reinforcement Learning (RL) [14]. RL uses samples generated through interaction with the environment in order to optimize performance. The key advantages of RL over classical planning approaches is its ability to handle large environments and when a model of the environment is not known or a solution using the model is not available due to it being computationally intractable [14].

For penetration testing, due to the complex nature of computer systems it becomes a major challenge to create and maintain an accurate model of how exploits will affect any given system. This is due to the constant evolution of attacks and the systems themselves. This property makes pentesting a good candidate for using RL, since it is possible to define penetration testing as an MDP but using RL we do not require the transition model. The main challenge facing RL is that it requires many interactions with the environment in order to learn an optimal policy. This feature has lead to many of the successful applications of RL being done in simulated or game environments where the RL agents is able to rapidly interact with its surroundings [15], [16], [27].

Presently, no freely available simulated environment exists for network penetration testing that can be used to train and test RL agents. This desire to apply RL to automated pentesting and the lack of training environment lead to the design and aims of this study. In the next chapter, we cover the design and implementation of a network attack simulator that can be used for training RL agents. Then in chapter 4 we cover the application of RL to penetration testing in more detail.

# Chapter 3

# The Network Attack Simulator

## 3.1 Introduction

The recent improvements to AI techniques have been greatly aided by the establishment of well-known, freely available performance benchmarks. These benchmarks can take many forms depending on the application domain, for example there is the arcade learning environment for testing generalised Reinforcement Learning algorithms [28] and the ImageNet dataset for computer vision [29]. These benchmarks become testing grounds, allowing researchers to compare the performance of algorithms with each other and over time. Currently, for network penetration testing there exists no light-weight, freely available benchmark that can be used for developing automated pentesting algorithms.

    For network pentesting, for it to be realistic, a benchmark would need to take the form of a network simulator that allowed agents to affect the network through scans and exploits. Presently there are numerous widely-used and freely available network traffic simulators, such as NS3 [30] and mininet [31]. These simulators are light-weight, capable of running effectively on a single machine, and able to emulate network topology and traffic with high-fidelity by using a very minimal virtualization of an actual OS. They do not, however, allow for modeling an attack on the network through launching exploits and gaining access to machines and so are not suitable for assessing penetration testing performance. Another option is to use a network of virtual machines (VM). This has the advantage of having high fidelity to the real world while also being flexible. The main disadvantage of using a VMs is the relatively high computational cost of running each VM, which can slow down the training of certain types of AI algorithms such as RL, and also the considerable computational power required to run larger networks. The best system currently available appears to be the Core Insight system [32]. This system operates the simulation at the OS level but only supports a subset of system calls, similar to the lightweight

network traffic simulators, in this way it is able to scale to networks of hundreds of machines on a single computer. However, this software is not open-source or free to the public and cost tens of thousands of dollars for a licence, which is out of the question for many researchers.

In this work, we design and build a new open-source network attack simulator (NAS). The NAS is designed to be easy to install with minimal dependencies and able to run fast on a single personal computer. It is also designed to model network pentesting at a higher level of abstraction in order to allow fast prototyping and testing of algorithms.

## 3.2 Design

The NAS is designed as two components. Firstly, the network, which contains all the data structures and logic for modeling traffic and attacks on the network. The second, is the environment which acts as the layer between the attacker and the network and models the attackers knowledge as an MDP.

### 3.2.1 The network model

The network model defines the organization, connection and configuration of machines on the network and is defined by the tuple *{subnetworks, topology, machines, services, firewalls}*. An example network made up of five subnetworks, 11 machines and firewalls between each subnetwork is shown in figure 3.2.1. This network could run any number of services as this is defined at the level of machine. We provide more details for each component in the following paragraphs. This model aims to abstract away some details of a real-world network that are not required when developing autonomous agents such as specific types of connections between machines and the location of switches and routers in the network. The reason for this abstraction is to try and keep the simulator as simple as possible and at the level of abstraction that the agent is expected to work at which is determining which scans or exploits to use against which machine and in what order. The specific details of performing each action, for example which port to communicate with, are details that can be handled by application specific implementations when moving towards higher fidelity systems. Penetration testing is already moving in this direction with frameworks such as Metasploit which abstract away exactly how an exploit is performed and simply provide a way to find if the exploit is applicable to the scenario and launch it, taking care of all the lower level details of the exploit [18]. This simpler network model is also used in order to keep it as general and easily scalable as possible.

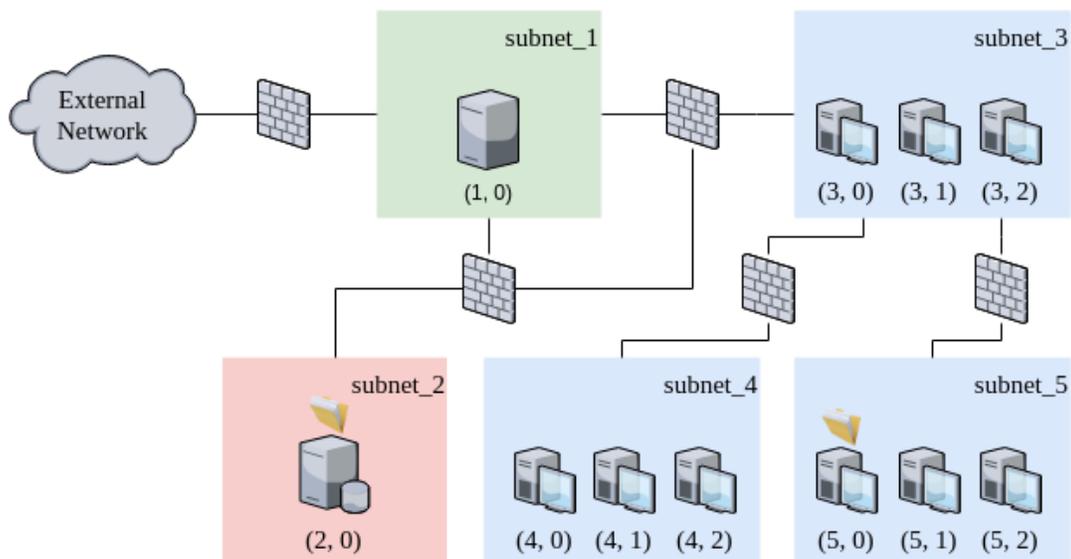

**Figure 3.2.1 | Example network** with five subnets, 11 machines and the sensitive documents located on machines (2, 0) and (5, 0).

## Subnetworks

Each network is made up of multiple sub-networks or subnets. A subnet is a smaller network within the larger network that is composed of a group of one or more machines that are all able to communicate fully with each other. Each subnet has its own subnet address, which is indicated as the first number in any machines address (e.g. the 4 in the address (4, 0)). This is a simplification of IP addresses which use a 32-bit string and a seperate 32-bit subnet mask to define the network, subnet and machine address. For the purpose of the NAS, it makes sense to use a simpler system since we are only dealing with a single network as opposed to IP addresses which deal with millions of machines on thousands of networks across the internet. Although all machines within a subnet can communicate fully, communication between machines on different subnets is restricted. Inter-subnet communication is controlled by the network topology and firewall settings.

## Topology

The network topology defines how the different subnets are connected and controls which subnets can communicate directly with each other and with the external network. As an example, in the network in figure 3.2.1 subnet 1 is the only network that is connected to the external world and subnets 1, 2 and 3 are all connected to each other while its only possible to communicate with machines on subnets 4 and 5 via a machine on subnet 3. In this way an attacker may have to navigate through machines on different subnets in order to be able reach the goal machines. We can view the network topology as an undirected graph with subnets as its vertices and connections as edges. As such we can represent it using an adjacency matrix, with rows and columns representing the different subnets, an example matrix is as shown in figure 3.2.2 for the example network in figure 3.2.1.

| subnet | 0 | 1 | 2 | 3 | 4 | 5 |
|---|---|---|---|---|---|---|
| 0 | 1 | 1 | 0 | 0 | 0 | 0 |
| 1 | 1 | 1 | 1 | 1 | 0 | 0 |
| 2 | 0 | 1 | 1 | 1 | 0 | 0 |
| 3 | 0 | 1 | 1 | 1 | 1 | 1 |
| 4 | 0 | 0 | 0 | 1 | 1 | 0 |
| 5 | 0 | 0 | 0 | 1 | 0 | 1 |

**Figure 3.2.2 | Example network topology** for the network in figure 3.2.1, represented using an adjacency matrix. Subnet 0 is used to represent the external network, while other subnets are represented by their ID number (i.e. 1 corresponds to subnet_1).

Machines

The most primitive building block of the network model is the machine. A machine in the NAS represents any device that may be connected to the network and hence be communicated with and exploited. Each machine is defined by its address, in the form of a (subnet_ID, machine_ID) tuple, it value and it's configuration. An example machine definition can be seen in figure 3.2.3. The value of a machine is defined by the user with higher values given to sensitive machines, that is machines that the attacker wants to gain access to or that the owner wants to protect. Each machine runs services that can be communicated with from other machines within the same subnet or on neighbouring subnets, firewall permitting. The services available on each machine define its configuration and each machine on the network will not necessarily have the same configuration. This is included since not every machine on the network will be the same as some can be expected to be used for different purposes E.g. Web servers, file storage, user machines. The services present on a machine also define its points of vulnerability, since the services are what the attacker is aiming to exploit.

```
Machine: {
    address: (1, 2),
    value: 0,
    configuration: {
        ftp: true,
        ssh: true,
        http: true,
    }
}
```

**Figure 3.2.3 | Example machine definition on the network.** The example is for the 2nd machine in subnet 1, which has no value (i.e. not one of the goal machines) and is running ftp, ssh and http services running that the attacker has an exploit for.

## Services

Services are used to represent any software running on a machine that communicates with the network. They are analogous to software that would be listening on an open port on a computer or connected device. Within the NAS services are considered to be the vulnerable points on any given machine, and can be thought of as services that have a known exploit which the attacker is aware of. In a real world scenario it would be the same as keeping track only of services that an attacker has a known exploit for, while ignoring any other non-vulnerable services. Based on this reasoning within the NAS, we assume each service is exploitable by one action, so the agents job is to find which service is running on a machine and select the correct exploit against it.

      Each service is defined by a unique ID, the probability its exploit will succeed and also the cost of using the exploit. Figure 3.2.3 shows an example machine in a network scenario where the attacker has exploits for the ftp, ssh and http services, while figure 3.2.4 shows the set of exploitable services and the associated success probability and cost of their exploits. The ID of each service can be any unique value and does not necessarily have to be a name related to a real world service. In this way it is easy for the NAS to generate test scenarios with any number of machines and services to aid in testing the scaling performance of agents by simply generating service IDs as needed. When investigating the application to more real world settings, the ID would be replaced with a specific service name and version, so it would be possible to track vulnerabilities and know what services require patching (e.g. Samba version 3.5.0).

```
Exploitable_services: {
    ftp: {
        probability: 0.8,
        cost: 3
    },
    ssh: {
        probability: 0.5,
        cost = 2
    },
    http: {
        probability: 0.2,
        cost: 1
    }
}
```

**Figure 3.2.4 | Example set of exploitable services on a network.** Each service definition contains a probability of its associated exploit working successfully and a cost for launching the exploit.

## Firewalls

The final component of the network model are the firewalls that exist along the connections between any subnets and also between the network and the external environment. Firewalls act to control which services can be communicated with on machines in a given subnet from any other connection point outside of the subnet. They function to allow certain services to be used and accessed from machines within a subnet with the correct permissions, while blocking access to that service from unwanted entry points. Each firewall is defined by a set of rules which dictate which service traffic is permitted for each direction along a connection between any two subnets or from the external network. Figure 3.2.5 shows an example firewall that sits between subnets 1 and 3 and which allows access to the ssh service on machines on subnet 3 from machines on subnet 1 and access to ftp and http services on machines on subnet 1 from machines on subnet 3. In a real world setting firewall rules are typically set by defining which port can be accessed, however for simplicity and since for most cases the same services are run on the same port numbers, we have decided to instead define rules by service rather than port.

```
Firewall_3: {                          Firewall_3: {
    connection : (1, 3)                    connection : (3, 1)
    permitted: {ssh}                       permitted: {ftp, http}
}                                      }
```

**Figure 3.2.5 | Example firewall on a network.** The example is for the firewall located on the connection between subnets 1 and 3 and defines which services are permitted in each direction.

## 3.2.2 The Environment

The environment component of the NAS is built on top of the network model and acts as the interface between the attacker and the network. It is responsible for modeling the attackers current knowledge and position during an attack on the network. For instance it tracks information about which machines the attacker has successfully compromised, which machines they can reach and what knowledge they have about services present on each machine. Its main function is to control the running of an attack from the beginning, where the attacker has not yet interacted with the network and has no information about any machine on the network, to the end of the episode where the attacker either gives up or is successful in compromising the goal machines on the network. We model the environment component of the NAS as an MDP since this framework is highly versatile and a building block used by many AI algorithms.

### MDP overview

The environment component models network pentesting problem as an MDP and as such is defined by the tuple $\{\mathcal{S}, \mathcal{A}, \mathcal{R}, \mathcal{T}\}$ (see chapter 2 for more details of MDP definitions). States are defined as the current knowledge and position of the attacker in the network. Actions are the available scans and exploits that the attacker can perform for each machine on network. The reward function is simply the value of any machines exploited minus the cost of actions performed. The transition function controls the result of any given action and takes into account action type, connectivity, firewalls and probabilistic nature of exploits.

### State space

A state, $s \in \mathcal{S}$, is defined as the collection of all known information for each machine on the network. That is, the state includes for each machine, if the machine is compromised or not, reachable or not and for each service, whether the service is present, absent or its existence is unknown. A machine is considered to be compromised if an exploit has successfully been used against it. While a machine is considered to be reachable if it is in a subnet that is publicly accessible (connected directly to external network), in the same subnet as a compromised machine or in a subnet directly connected to a subnet that contains a compromised machine. The state space is therefore all possible combinations of compromised, reachable and service knowledge for each service and for each machine. Hence, the state space grows exponentially with the number of machines and services on the network. Equation (3.1), shows the size property of the state space, $|\mathcal{S}|$, where $|E|$ is the number of exploitable services and $|M|$ is the number of machines in the network. The base of for the exponential is 3, since for each exploitable service the agents knowledge can have one of three values: *present*, *absent* or *unknown*.

$$|\mathcal{S}| \in O(\,3^{|E||M|}) \qquad (3.1)$$

An example network and an associated state are shown in figure 3.2.6. In this state the attacker has successfully compromised the machine at address *(1, 1)* and so can now communicate with machines on subnets 2 and 3, which is indicated by reachable being set to *true* for each machine on those subnets. Additionally, the configuration of the machine at *(3, 1)* is known which would have been gained through a scan action, while the configuration for machine *(2, 1)* is *unknown*. Note that the state does not include any information about the firewall settings, since this would require privileged access to determine. For this simulator we assume it is not possible for the attacker to gain this information and it must instead learn it indirectly through the success and failure of exploit actions.

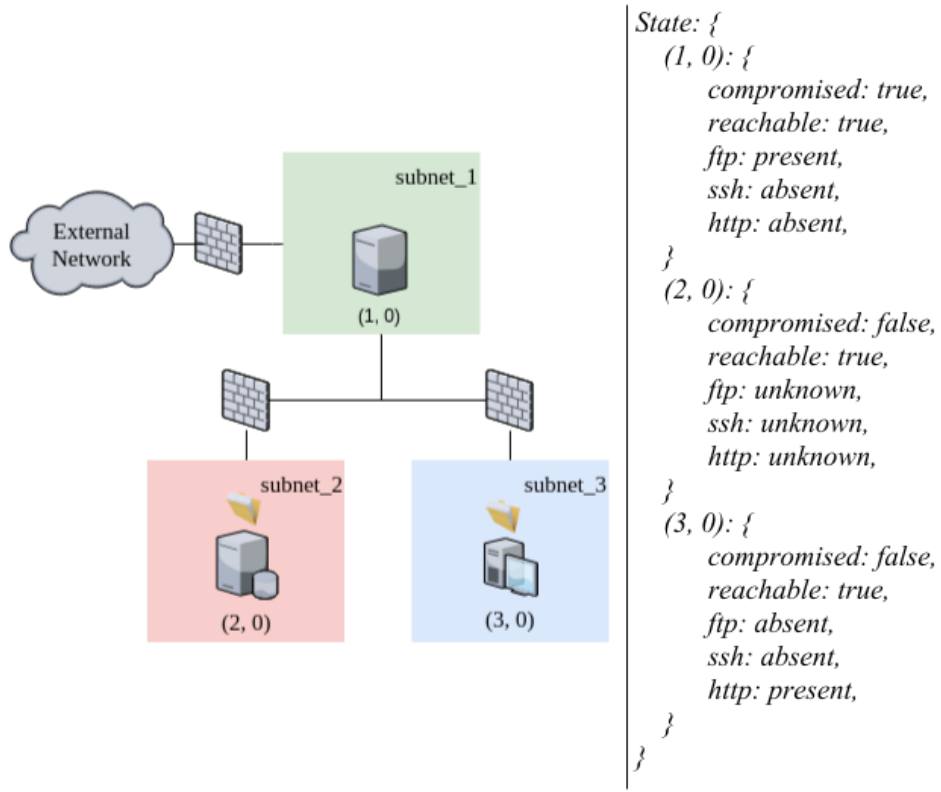

**Figure 3.2.6 | Example network and state**, where the first machine in the network has been compromised and machine (3, 0) has been scanned to get service information. No information has been gathered regarding services on machine (2, 0)

### Action space

The action space, *A,* is the set of available actions within the NAS and includes a single scan action and an exploit for each service and each machine on the network. The scan action is designed to mimic the Nmap complete scan, which returns information about which services are running on each port of a given machine and also versions of each service [19]. In reality more targeted scanning may be required to discover complete information about specific services, but for many use cases, Nmap scans return the information required to determine which service is running. Scan actions are considered to be deterministic, always returning information about the presence or absence of a service.

For each possible service on the network, there is a matching exploit action. Each exploit action can be deterministic or non-deterministic depending on the configuration of the environment chosen by the user. A successful exploit action will result in the target machine becoming compromised. The success of any exploit is determined by whether the target machine is reachable, the target service is present, if that service is blocked by the firewall or not and also the success probability of the action.

Each action also has an associated cost which can be set when configuring the environment. This cost can be used to represent any metric such as the time, skill, monetary cost or noise generated for a given action, depending on what performance metric is trying to be optimized for. Figure 3.2.7, shows example scan and exploit action definitions. Both actions target the same machine at address *(1, 0)*. "Action_1" is a scan with cost 1 while "Action_2" is an exploit for the ssh service and has a cost of 3 and a probability of success of 0.8.

```
Action_1: {                    Action_2: {
    target: (1, 0),                target: (1, 0)
    type: scan,                    type: exploit
    cost: 1                        service: ssh
}                                  cost: 3
                                   probability: 0.8
                               }
```

**Figure 3.2.7 | Example scan and exploit actions** against machine (1, 0). Exploit is targeting the ssh service.

### Reward

The reward function is used to define the goals of the autonomous agent and what is trying to be optimized by the agent. The reward is defined over a transition $\mathcal{R}(s, a, s')$, so starting from one state *s* taking action *a* and ending in the resulting state *s'* (eq. 3.2). The reward for any transition is equal to the value of any newly compromised machine in the next state *s'* minus the cost of action *a*. So if no machine was compromised, then the reward is simply the cost of the action performed. With this reward function, the goal of the attacker becomes to try and compromise all machines with positive value on the network while minimizing the number or cost of actions used. This mimics a real world nefarious attacker, whos goal we assume is to to retrieve privileged information or gain privileged access on the system. Using the NAS it is possible to set these goals by changing the value of certain machines on the network, so for example machines that may contain sensitive documents or contain privileged control on the network.

$$\mathcal{R}(s', a, s) = value(s', s) - cost(a) \qquad (3.2)$$

Where, *value(s', s)* returns the value of any newly compromised machines in *s'* from *s* or 0 if no new machines were compromised and *cost(a)* returns the cost of action *a*.

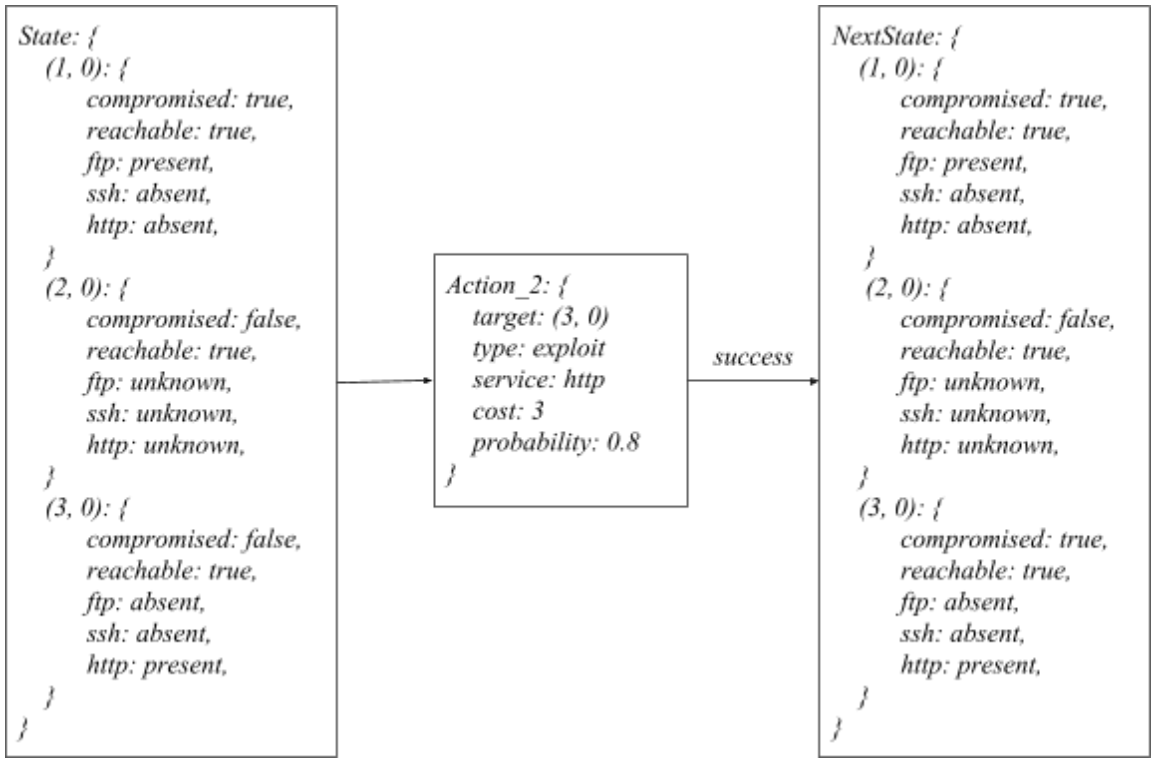

**Figure 3.2.8 | Example state transition** following a successful exploit launched against the http service on machine (3, 0).

Transition function

The transition function, $\mathcal{T}$, determines how the environment evolves over time as actions are performed. For the NAS the next state depends on whether an action was successful or not, which in turn depends on a number of factors. Specifically, whether the action target is reachable, whether the action is a scan or exploit, if traffic is allowed for the target service between a compromised machine and the target (i.e. if target is on seperate subnet), whether the target machine is running target service and finally, for non-deterministic exploits, the success probability of the exploit being used. Figure 3.2.8 shows an example transition, where an attacker successfully exploits the *http* service on the machine at address *(3, 0)*. The new state represents this success through machine *(3, 0)* having its compromised flag set to *true*.

## 3.3 Implementation

The goals of the NAS are to be fast, easy to install and able to be used for fast prototyping of AI agents. To help meet these goals the program was written entirely using the Python programming language (Python Software Foundation, www.python.org) and popular well-supported open source libraries. Specifically, the libraries used were numPy for fast array based computation and Matplotlib and NetworkX for rendering [33]–[35]. It would have been possible to build a faster simulator using a lower-level language such as C++, however another reason to use Python was it's popularity for machine learning research and its well-supported deep learning libraries such as Tensorflow [36] and Keras [37] that are commonly used for developing reinforcement learning agents such as the one used in the next chapter of this thesis. Please see Appendix A for details on and for access to source code for this project.

      A diagram of the NAS architecture is shown in figure 3.3.1. There are a number of different modules which handle the network model, MDP and other functions with the main component being the Environment module. The Environment module is the main interaction point for an agent and has four main functions: load, reset, step and render.

      The load function loads a new environment scenario either by generating a network from a standard formula or loading a network from a configuration file (more details of each are provided in the following section).

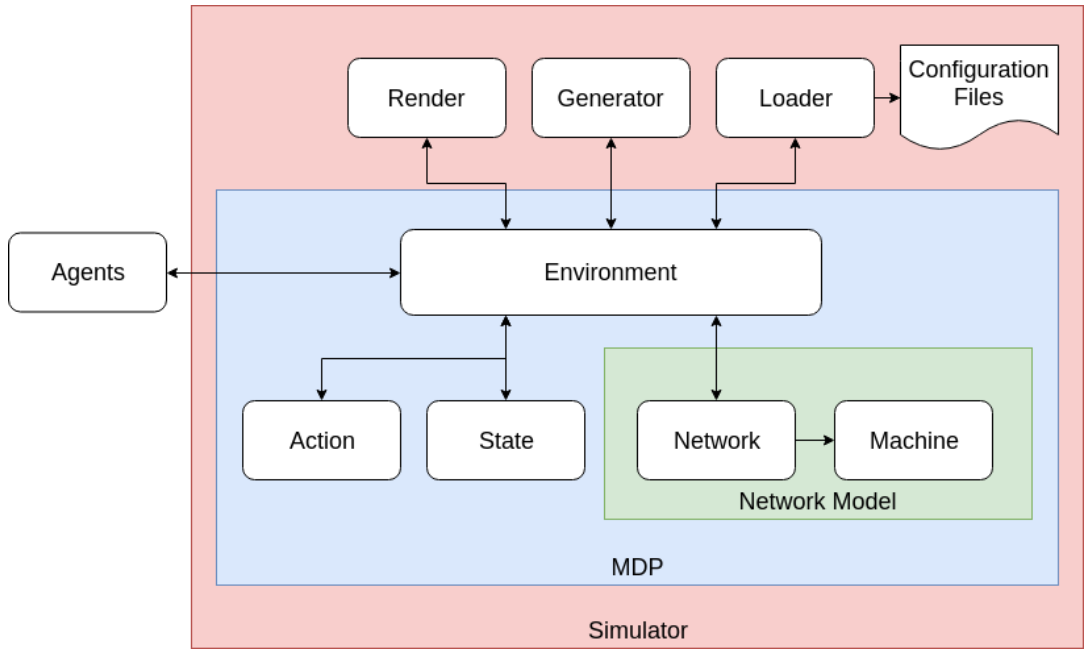

**Figure 3.3.1 | Network Attack Simulator program architecture**

    The reset function sets the environment to its initial state and returns the starting state to the agent and is synonymous with starting a new attack on the network. The initial state is where the agent has not compromised any machines on the network, only networks connected to external network are reachable and no information is known about services running on any machines on the network. An example network and it's associated initial state is shown in figure 3.3.2.

    The step function is the main interaction point for an agent with the environment. It takes an action and performs the transition given the the current state of the simulator and returns the next state, reward for performing the action and whether the goal has been reached. A typical cycle of training or testing of an autonomous agent involves: i) resetting the NAS to get the start state, ii) using the state to choose an action, iii) executing the action against the environment using the step function to receiving the next state and reward, iv) repeat steps (ii) and (iii) until the goal is reached or the time limit expires. This complete cycle would be a single episode and an agent may then reset the environment and repeat this for as many episodes as desired.

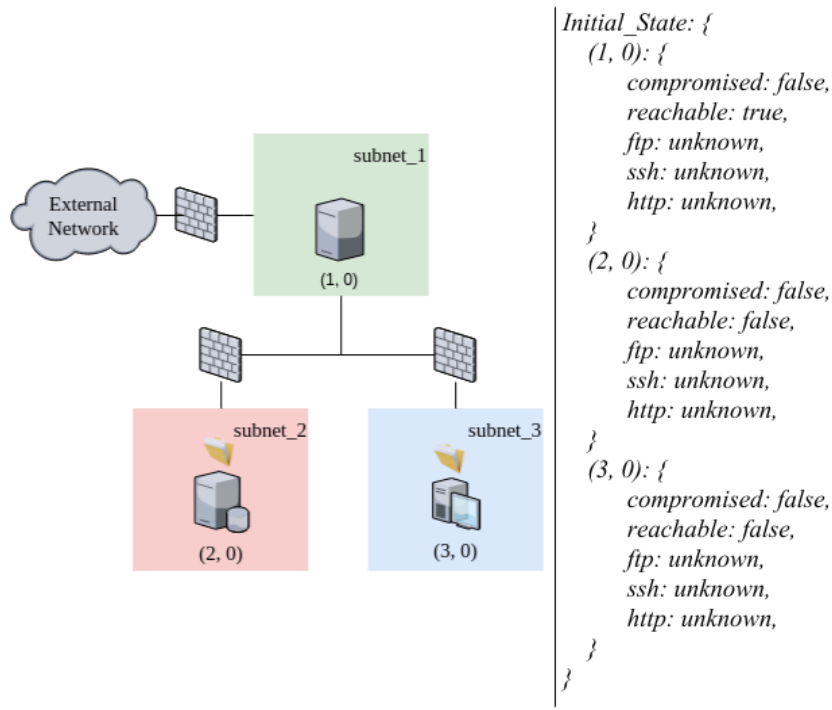

**Figure 3.3.2 | Example network and initial state**, where the network contains three exploitable services: ftp, ssh and http.

 

      The render function provides a number of ways to visualize the environment and network and also an attack episode on the network. The first option is to simply render the network as a graph, where each vertices is a machine in the network and machines on the same subnet are grouped together, while edges are connections between each machine within a subnet and between subnets (fig. 3.3.3). This option allows the user to visualize the topology of the network. The second option allows the user to visualize how the state of the network changes over the course of an attack episode (fig. 3.3.4). This mode shows the state of the network, along with the action and reward received for each step and can be used to visualize the attack policy of an agent and identify exploited services along the attack path.

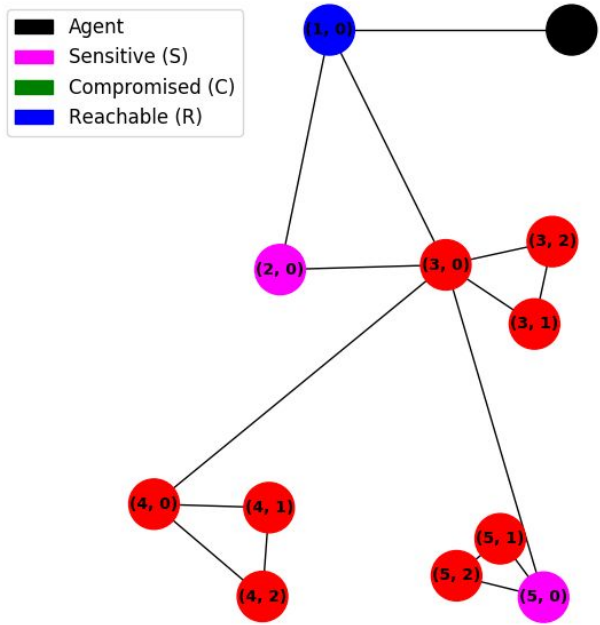

**Figure 3.3.3 | Simulator rendering output of example network in figure 3.2.1.** Each node represents a machine on the network, which machines clustered by subnet. Each edge between nodes with the same subnet ID represents connectivity within a subnet, while edges between nodes with different subnet ID represent inter-subnet connectivity. Pink nodes are the goal machines that contain "sensitive documents", red and pink nodes both represent machines that are not reachable by agent, blue nodes represent machines reachable by agent and the black node represents the position of the agent.

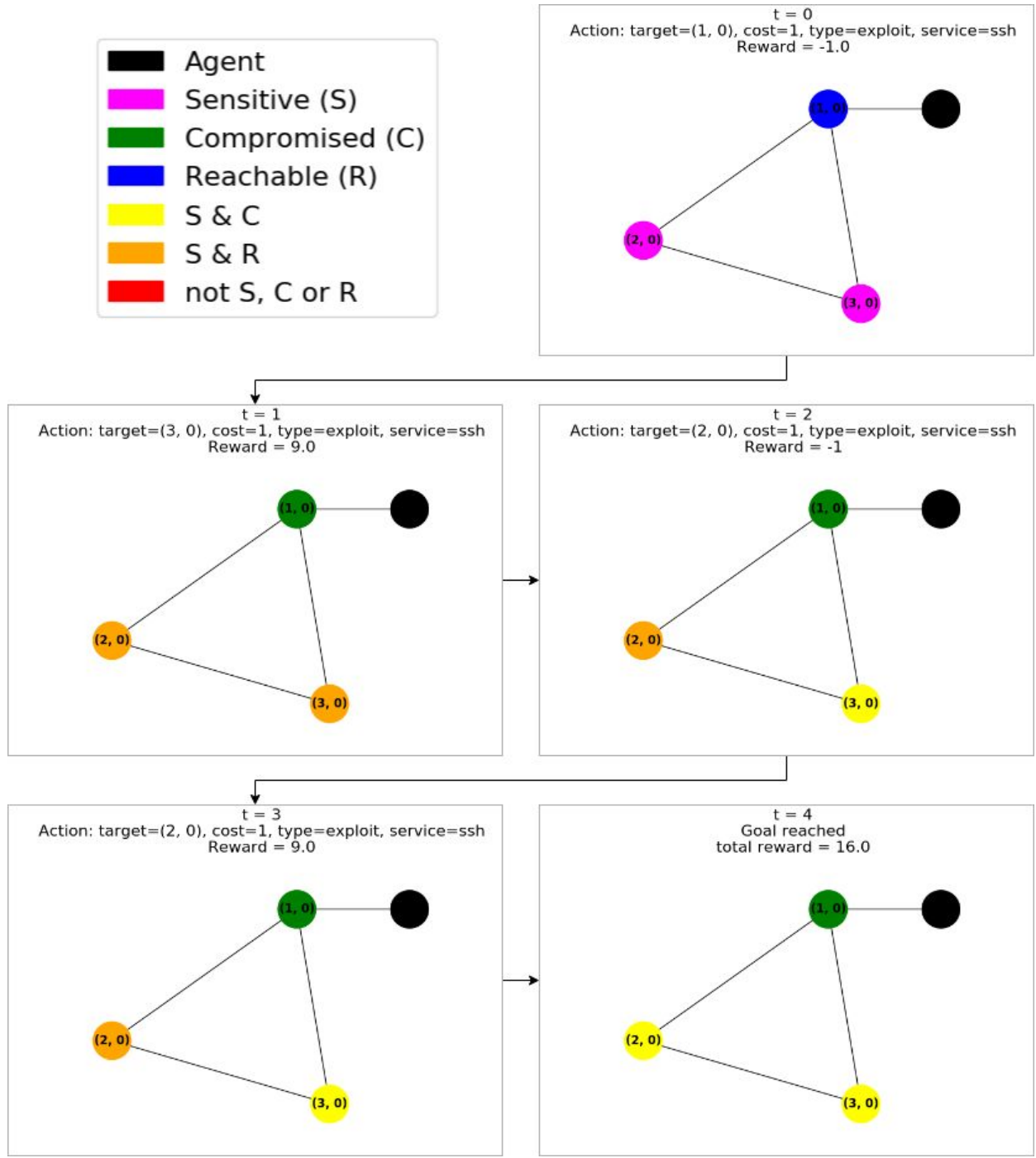

**Figure 3.3.4 | An example episode rendered using the Network Attack Simulator.** The episode shows each timestep from start (t = 0) till the end of the attack (t = 4), showing the action performed and reward received. The different node colours represent changing states of machines on the network.

Configuring a new environment

The simulator is capable of running any arbitrary network structure as defined by the user in a configuration file. A custom network is defined by the tuple: {s*ubnets, topology, sensitive machines, services, service exploits, machine configurations, firewalls*}. Table 3.3.1 provides a description of each required parameter and figure 3.3.5 shows an example file. The configuration files are written using the YAML ain't markup language (YAML, http://yaml.org/) since it is easy to read, write and well supported by Python.

Table 3.3.1 | Description of parameters required for defining a custom network

| Parameter | Description |
|---|---|
| subnets | The number and size of each subnet |
| topology | Connectivity of each subnet defined using an adjacency matrix |
| sensitive machines | The address and value of each sensitive machine on the network |
| services | The number of possible services running on any given machine |
| service exploits | The cost and success probability of each service exploit |
| machine configurations | Which services are running on each machine in the network |
| firewalls | Which service traffic is permitted along each subnet connection on network |

Generating an environment

In order to allow for rapid testing of agents on different sized networks and include a standard network topology that can be utilized as a benchmark for researchers to compare agents performance, we included the option to automatically generate a network for a given number of machines, *M*, and services, *E*. The generated network is based on the network description first developed by Sarraute et al [12] for testing the performance of POMDP agents and which was subsequently used by Backes et al [38].

The network is divided into three main subnetworks: i) the demilitarized zone (DMZ), ii) the sensitive subnet and the iii) user subnetworks (fig. 3.3.6). The *M* machines are divided into each subnet as follows, one machine in each of the DMZ and sensitive subnets and the remaining *M* - 2 machines in the user subnetworks which are connected in a binary tree structure with a max of five machines per user subnet and connections only between parent and child nodes in the tree. The DMZ, sensitive and root user subnets are all connected, and the DMZ is connected to the external network. There are network contains two sensitive machines to reach; one located in the sensitive subnet and the other on in a leaf subnet of the user tree.

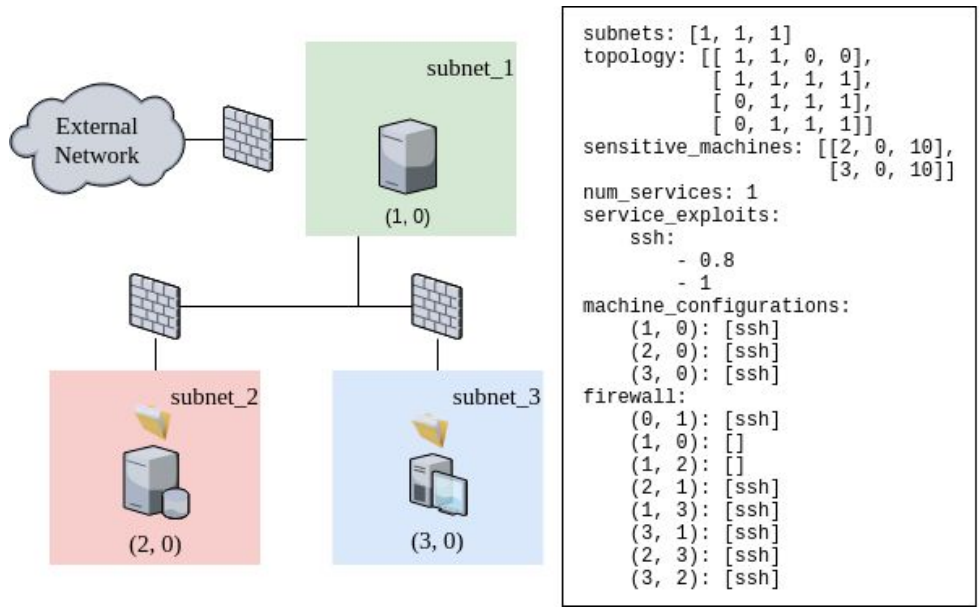

**Figure 3.3.5 | Example network configuration file and generated network topology.** The file is defined using YAML

    The distribution of configurations of each machine in the network are generated using a Nested Dirichlet Process, so that across the network machines will have correlated configurations (i.e. certain services/configurations will be more common across machines on the network) [38]. This is used in order to more accurately model the correlation of machine configurations seen in real-world networks, where most machines on a network will be running the same services.

    The success probabilities of each exploit can be chosen, however, the default is to randomly sample probabilities from a distribution based on the attack complexity score distribution of the top 10 most common exploits used in 2017 [39]. The attack complexity score is a metric generated by the Common Vulnerability Scoring System (CVSS) and is used to reflect how hard it is to find an exploitable machine along with the success probability and the skill required to use a given exploit [40]. Specifically, the probabilities were chosen based on the attack complexity score for CVSSv2, which scores exploits as having either 'low', 'medium' or 'high' attack complexity [41]. We take the same approach as Backes et al [38] and set probability of success to be 0.2, 0.5 and 0.8 for 'low', 'medium' or 'high' attack complexity

respectively. Using this approach we hope to try and model the distribution of exploit success probabilities found in the real world.

Finally, firewalls exist between each subnet on the network. The firewalls between user subnets permit all traffic while the permitted services for each firewall along connections between the DMZ, sensitive and root user subnets are chosen at random, but in such a way that there is always at least one service that can be exploited on at least one machine for any given subnet from any connected subnet. Permitted services are similarly chosen for the connection between the external network and the DMZ. Additionally, it possible to set the restrictiveness of the firewalls which limits the max number of services that are allowed for any given firewall so a user can model a more or less regulated network.

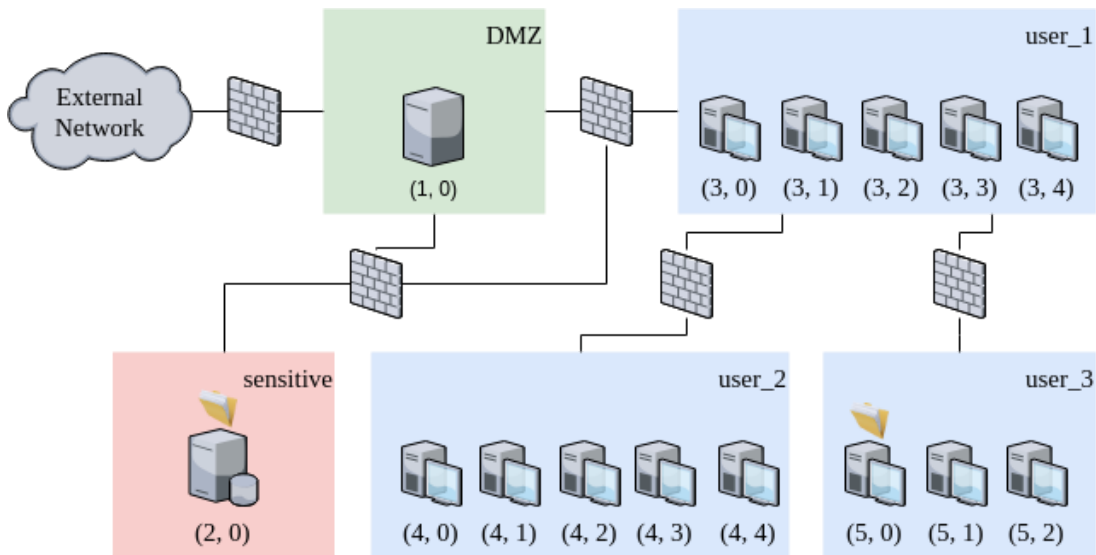

**Figure 3.3.6 | Example generated network with M = 15**

## 3.4 Results

In this section we provide the results of some experiments run on the NAS to test its performance and scaling properties. The metrics used were actions per second and load time and we measured these versus the number of machines and services on the network. All experiments were conducted using a single core on a personal laptop computer running the Linux Ubuntu 18.04 operating system. The test machine was running an Intel Core i7 7th Gen CPU at 2.7 GHz and had 16GB of RAM.

Load time of the NAS was measure for a range of machine from 10 to 1000 and a range of services from 10 to 1000. Figure 3.4.1 shows the mean load time for the NAS versus the number of machines and services. For mean load time, the minimum time was 0.0007 ± 0.00007 sec (mean ± standard deviation) found in the smallest simulator size tested with 10 machines and 10 services and the maximum time was was 3.557 ± 0.06 sec and was for the largest simulator size of 1000 machines and 1000 services.

For comparison the load time of a single VM was also measured. We used Oracle VM VirtualBox virtualization software [42] running the linux based Metasploitable 2 operating system [43] as the test VM. Mean load time for a single machine with no GUI (headless mode) was 0.249 ± 0.023 sec, averaged over 10 runs. This load time is over 300 times larger than loading 10 machines and services in the NAS.

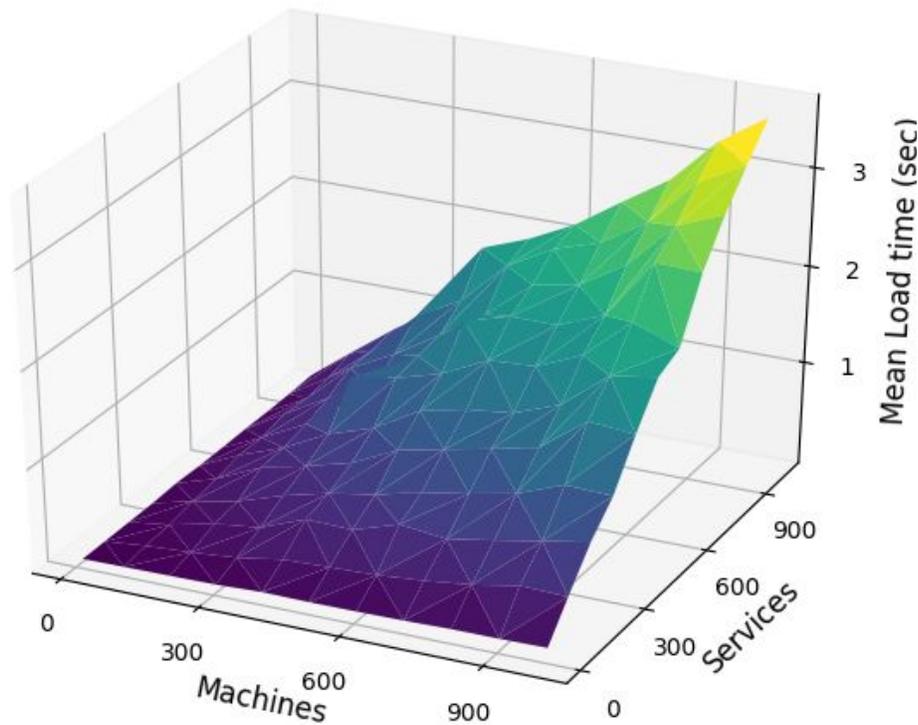

**Figure 3.4.1 | Network Attack Simulator load times versus number of machines and services.** Times were the average of 10 different runs for each tested pair of machines and services, using different random seeds.

To test the scaling properties of the simulator during regular use we measured the number of actions per second versus NAS size. We ran experiments on simulators using a range of machines (10 to 480) and services (10 to 480) and measured the actions per second and averaged this over a number of runs (fig. 3.4.2). For the mean actions per second of the settings tested, the minimum was 17329 ± 1907 actions per second for the simulator with 480 machines and 10 services, while the maximum was 126383 ± 3843 actions per second for the simulator with 10 machines and 30 services.

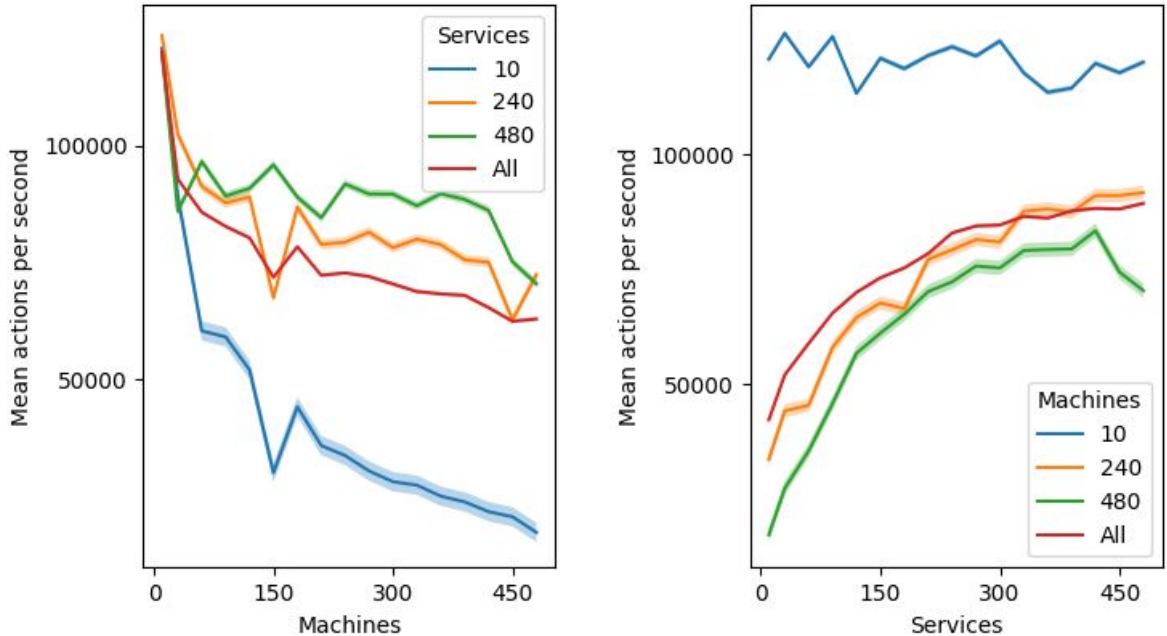

**Figure 3.4.2 | Scaling performance of Network Attack Simulator** in terms of mean actions per second versus the number of machines and services. Results are averaged over 10 runs with 10,000 actions per run, using deterministic actions. Actions performed are selected from action space in sequence, to try and keep consistency across runs. The left figure shows performance versus number of machines for networks running 10, 240 and 480 services. The right figure shows performance versus number of services for networks with 10, 240 and 480 machines. In both figure the red 'All' line is averaged over all values of services (left) or machines (right) tested.

As a comparison, we measured the mean time required to perform a scan and an exploit between two VMs. We used one attacker VM running Kali Linux (https://www.kali.org/) and one vulnerable VM running the Metasploitable OS. We used a standard Nmap scan of a single port, specifically port 21 which was running the ftp service, while the exploit used was for a backdoor in version 2.3.4 of the VSFTPD ftp server (https://www.exploit-db.com/exploits/17491/) run using the Metasploit framework. The mean time required for the Nmap scan was 0.253 ± 0.03 sec, averaged over 10 executions. This is equivalent to approximately 4 actions per second. It was not possible to get an exact measure of time required for performing the exploit, as it was being used from within the Metasploit framework, however based on external timing (using a stopwatch) the time required from when the exploit was launched until a shell was opened on the vulnerable machine was between 3 and 5 seconds, or roughly 0.2 to 0.3 actions per second.

## 3.5 Discussion

The aim for this part of the study was to design and build a fast and easy to install simulator for automated pentesting. The implementation of the NAS which only requires the Python programming language and common open source libraries is easy to deploy and fast, when compared with using VMs. In this section, we further discuss the performance, advantages and disadvantages of the simulator in context with other options that currently exist.

    We measured the performance of the simulator in regards to two metrics: load time and time taken to perform an action (i.e. actions per second). In terms of load time, the average time required increased linearly with the number of machines and number of services present in the network (fig 3.4.1). In terms of practical use, for worst case where there was 1000 machines and 1000 services the load time was approximately 3.5 sec which is many orders of magnitude lower than the time that an agent would spend training on the environment and so would not be a bottleneck in terms of time. Additionally, when compared with the load time of a single VM, which was roughly 0.25 seconds to launch the VM and would require additional time for the OS to load, the NAS is significantly faster. The benefits of this rapid load time, means it is possible to gather more training data faster which will speed up training time for interactive agents such as RL.

    The time required to perform an action was measured using actions per second (fig 3.4.2). Performance decreased as the number of machines in the network increased, which is expected due to the state space growing with the number of machines and the simulator having to check more machines and also generate a larger state as the number of machines grows. Conversely, performance improves with the number of services when the number of machines in the network is larger. After further investigation this is mainly due to there being a much smaller number of successful exploit actions when the number of services increases, with successful exploits requiring an update to the state and so requiring more time to process. Based off of this the performance expected would be less than reported assuming the agents are performing a higher proportion of successful actions. In practical terms, the worst case performance of the simulator sizes tested was around 17,000 actions per second for a network with 480 machines and running 10 services, which is orders of magnitude faster than using a VM network which performed could perform roughly 4 Nmap scan actions and 0.3 exploit actions per second for a simple network of two VMs. For AI algorithms which rely on interaction with the environment for learning an attack path, such as RL, the speed in which these interactions occur has significant impact on how fast they are able to generate a useful attack plan. For these algorithms

the speed of the NAS compared with what might be expected from a VM will be a great benefit when designing these algorithms.

Currently, there do exist fast higher fidelity network simulators that make use of lightweight virtualizations. Most of these simulators are currently applicable only for modelling network traffic (e.g. NS3 [30] and mininet [31]). In terms of simulating pentesting, the Core Insight simulator [32] is a great example of what is possible, offering a higher fidelity lightweight virtualized network system. As mentioned previously, this system however requires a commercial licence and quite expensive. But even ignoring the price tag, the higher fidelity also means more information is required to configure the simulator, so there is some loss in versatility. One advantage of the NAS over these kinds of lightweight virtualizations is its ability to model arbitrary services and exploits and generate networks of any size. This comparison between higher fidelity simulator and the more abstract simpler NAS presents a trade-off between fidelity and versatility. Obviously, higher fidelity is key when it comes to applying technologies to a real-world setting, however during the early stage of development it is useful to have a versatile simulator especially when designing interactive AI agents.

## 3.5 Conclusion

Overall, the presented NAS offers an easy to deploy, fast and flexible penetration testing testbed for developing automated agents. The main weakness of this approach is it's fidelity to a real environment which contains much more complexity in terms of the exploits, services, machine configurations and network dynamics, however it offers an abstract environment that has a number of use cases. Specifically, for rapid prototyping of agent design, in terms of representing state, action and reward and handling and learning of the world dynamics. Also for investigating the performance properties of algorithms as network topology and size changes. As a next step it will be necessary to develop higher fidelity systems to further test and develop autonomous agents before they can be used in real world applications.

# Chapter 4

# Penetration Testing using Reinforcement Learning

## 4.1 Introduction

The second part of this research aims to investigate the use of Reinforcement Learning (RL) for automated pentesting. RL is a field of study as well as a class of solution methods for learning the mapping from states to actions in order to maximize a reward [14]. In RL there is an environment and an agent who interacts with the environment with the aim of learning the optimal actions to take from each state. There are four main components of a RL system, other than the environment and agent, these are the policy $\pi$, the reward function, $\mathcal{R}(s', a, s)$, the value function, $V(s)$, and optionally the model of the environment, $\mathcal{T}$ [14]. The policy is a mapping from the state space, $\mathcal{S}$, to the action space, $\mathcal{A}$. We want the agent to find the optimal policy $\pi^*$ that chooses the action, $a$, from any state, $s$, that maximizes the total expected discounted reward. The reward defines the immediate reward for the current state and is sent by the environment on each time step. The value function specifies the value of a state over the long run, such that the value of a state, $s$, (i.e. $V(s)$) is the total accumulated reward the agent can expect to receive in the long term starting from that state. The model of an environment is something that tells the agent something about how the environment will behave and allows the agent to make inferences. The transition model, $\mathcal{T}(s', s, a)$ is an example of a model, since it gives the agent information about the expected future state given the current state and chosen action. In RL when a model is present it is know as *model-based* RL, while if no model is present then it is *model-free* RL. Model-based problems are typically solved using planning while model-free problems must rely on trial-and-error in order to find the optimal policy for the environment.

So far there has been no published applications of RL to automated penetration testing. One of the main advantages of using a RL approach is that it allows us to approach the problem with no assumed prior knowledge, or model, about the action outcome probabilities for any given state and instead allows these to be learned by the agent. This provides a solution to one of the challenges of automated pentesting which is producing and maintaining an up-to-date accurate model of the exploit outcomes. The fast evolving and diverse nature of software systems and exploits means in order to produce an accurate model it would be necessary to test any exploit on a wide range of systems and repeat this process over time. Using RL on the other hand, requires only a definition the state representation, the set of actions an agent can take and a reward function. The agent then explicitly learns the model of the environment through interaction. This means that as the cyber security space evolves it would only be necessary to update the actions that an agent can take and leave the modelling to the agent.

Over the years many different RL algorithms have been developed [14]. For this study we will be using Q-learning which is a RL algorithm for learning an optimal policy in a model-free problem [44]. It relies on using experience to learn the Q-value function, *Q(s, a)*, which tells the agent the expected reward if they perform action *a* from state *s*. It has been proven that given enough time and exploration of the environment this algorithm will converge on the optimal Q-values for each state and action pair. Once the Q-value function converges we can then use it to determine the optimal action for a state simply by choosing the action with the highest Q-value and hence use it to find the optimal policy for a given environment.

RL is a powerful and versatile approach for solving MDPs however it can be harder to implement and performance can be variable depending on the rate of convergence of the value function. This difficulty may be one of the key reasons it has not yet been utilized in automated pentesting. This study aims to investigate the applicability of RL to the penetration testing domain. We do this by first modelling penetration testing as an MDP where the transition model $\mathcal{T}$ is unknown and then using RL in a simulated environment to generate attack policies.

In the following section, 4.2, we provide some more details about RL as it applies to this study, then in section 4.3 we frame automated pentesting as an RL problem using a MDP. In section 4.4 and 4.5 we use the network attack simulator (NAS) we developed to test the capabilities of RL in terms of whether it is capable of finding an attack path in an network environment, how optimal is the attack path, scaling performance and generality. Finally, section 4.6 provides a discussion of the experimental results and section 4.7 provides some concluding remarks.

## 4.2 Background

RL algorithms learn optimal policies through interaction with the environment. This done by starting from some initial, typically random, policy then iteratively learning the values of taking a certain action for a given state, $Q(s, a)$, by choosing an action based on the current policy, applying that action to the environment, then updating the state-action value, $Q(s, a)$, based on the received experience (fig. 4.2.1) [14]. Specific RL algorithms differ based on how they choose actions, update their value estimates for the value function and the form of the value function.

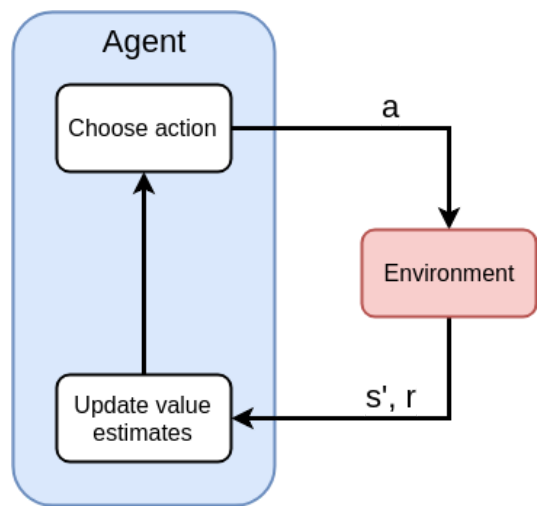

**Figure 4.2.1 | The Reinforcement learning cycle.**

There are a number of different action selection strategies available but two of the most commonly used, and the ones used for this study, are ε-greedy and upper confidence bound (UCB) action selection [14], [45]. Both these strategies aim to balance exploration versus exploitation. The ε-greedy action selection strategy does this by choosing a random action with ε probability, and choosing the current best action the rest of the time (eq. 4.1). This forces the agent to randomly explore actions that it does not think are valuable given the current experience. It is common to implement ε-greedy along with ε-decay which decreases the value of ε over time so that as the agents estimates of action values improve it chooses random actions less frequently. UCB action selection, on the other hand, uses an extra exploration term when doing action selection (eq. 4.2). This extra term increases the value of actions that have been taken less frequently and acts to measure the uncertainty of action value estimates. For this study we implement algorithms that make use of both these action selection methods in order to have a comparison and investigate how action selection might affect the use of RL in automated pentesting.

$$a_t = \begin{cases} \underset{a \in A}{argmax}\ Q(a) & with\ p(1-\varepsilon) \\ random\ a \in A & with\ p(\varepsilon) \end{cases} \qquad (4.1)$$

$$a_t = \underset{a \in A}{argmax}\left[Q(a) + c\sqrt{\frac{\ln t}{N_t(a)}}\right] \qquad (4.2)$$

We make use of Q-learning in our RL implementations for the value update strategy. Q-learning is an off-policy temporal-difference algorithm for learning the action-state values and is defined by the update function in equation (4.3) [44]. Where $\alpha$ is the step size, which controls how much to move current estimate towards the new estimate and $\gamma$ is the discount factor which controls how much to weight immediate versus future rewards. Q-learning has been shown to converge to the optimal state-action values, as the number of visits to state-action pair approaches $\infty$.

$$Q(S_t, A_t) \leftarrow Q(S_t, A_t) + \alpha\left[R_{t+1} + \gamma max_a Q(S_{t+1}, a) - Q(S_t, A_t)\right] \qquad (4.3)$$

The final key aspect of RL algorithms that differs between implementations is what form the value function, *Q(s, a),* takes. There are two main options: i) *tabular* and ii) *function approximation [14]*. Tabular methods use a table like data-structure to store the state-action value for each state-action pair with pairs being updated as the agent receives experience. Function approximation methods, on the other hand, use a function to generate the state-action values and update the parameters of the function to improve these estimates as the agent gains more experience. In this study we implement both tabular and function approximation methods to investigate how both these methods perform when applied to network pentesting.

Tabular methods have the advantage that they are simple to implement and can often find exact optimal solutions. Their use, however, is limited to problems with relatively small state sizes since each state-action pair must be stored. They also treat each state independently and so cannot make use of knowledge learned about different but similar states to generalize to unseen states.

Function approximation methods on the other hand can be used with arbitrarily large state spaces and are capable to generate to unseen states. However, they are more complicated to implement, requiring the use of an extra function to approximate the value function with performance strongly affected by the function representation used. There are many options for how to represent the function, however the method gaining most attention recently and for which the biggest RL improvements have been seen in recent years has been the using of deep neural networks [15], [46]. The combination of using Q-learning and neural networks for function approximation is known as Deep Q-learning and has been used to produce state of the art results in a number of environments including the game of Go and on many Atari video games [15], [46]. As such it has been shown to be a powerful algorithm when dealing with large state spaces.

## 4.3 Penetration testing using Reinforcement learning

In order to use model-free RL for Automated Pentesting we need to frame the problem as an MDP leaving the model unknown. As discussed previously, an MDP is defined by the tuple {$\mathcal{S}$, $\mathcal{A}$, $\mathcal{R}$, $\mathcal{T}$}. We represent states, actions and reward as presented in Chapter 3, with states being the status and configuration knowledge of the agent for each machine on the network, actions as the available scans and exploits for each machine and reward given by the value of newly compromised machines minus the cost of action (fig. 4.2.1). The transition model, $\mathcal{T}$, is of course unknown.

**Table 4.2.1 | Reinforcement learning MDP definition,** where |M| represents the number of machines on the network and |E| represents the number of exploitable services.

| Component | Definition |
|---|---|
| $\mathcal{S}$ | \|M\| x {compromised} x {reachable} x \|E\| x {machine service knowledge} <br><br> *Where:* <br>    *compromised ∈ {true, false}* <br>    *reachable ∈ {true, false}* <br>    *Machine service knowledge ∈ {absent, present, unknown}* |
| $\mathcal{A}$ | \|M\| x {scan, exploit} x \|E\| |
| $\mathcal{R}$ (s', a, s) | value(s', s) - cost(a) |
| $\mathcal{T}$ (s', a, s) | unknown |

Since this is a primary investigation into the use of RL for pentesting, we use one key assumption: the agent has complete knowledge of the network topology. This means the agent knows the address of each machine on the network and their connectivity. This assumption is made in most attack planning approaches, and is based on the availability of the topology data from a client requesting a penetration testing assessment [13]. This assumption could be relaxed in principle, but would mean the state representation vector would change over time since we would not know beforehand how many machines are on the network and so the size of a state would grow as more machines are discovered. Applying RL to problems where the state representation is dynamic is much more difficult and the vast majority of work into RL involves stationary state representations. For this reason we believe it is better to have this assumption at this early stage of the research and then if RL proves promising, perhaps attempt to apply it with no knowledge of the network topology and mimic exactly the point of view of a real attacker.

## Reinforcement learning algorithms

As there are currently no studies investigation the use of RL for automated pentesting we decided to implement a range of RL algorithms. Specifically, we make use of three different Q-learning algorithms: tabular Q-learning using ε-greedy action selection (tabular ε-greedy), tabular Q-learning using UCB action selection (tabular UCB) and deep Q-learning using a single layer neural network and ε-greedy action selection (DQL). These algorithms were chosen as they provide both tabular and function approximation RL implementations, while using tabular UCB also provides the opportunity to investigate how action selection strategy affects performance in the pentesting domain.

The algorithms implemented for tabular ε-greedy and tabular UCB are presented in Algorithm 1 and Algorithm 2, respectively. These are standard implementations of Q-learning based on work presented in Sutton and Barto [14]. The action-value function is stored using a hashmap with states as keys and action-values as values. The key difference between the two tabular implementations is the action selection method along with the use of an extra data structure to store visit counts for each state-action pair in the tabular UCB implementation. Otherwise the algorithms are the same.

**Algorithm 1** Tabular Q-learning with $\epsilon$-greedy action selection

1: Initialize $Q(s, a)$, for all $s \in \mathcal{S}, a \in \mathcal{A}(s)$, arbitrarily
2: **for** episode = 1, V **do**:
3:     $s_1$ = initial state
4:     **for** step = 1, T **do**:
5:         With probability $\epsilon$ select random action $a_t$
6:         otherwise select $a_t = \arg\max_a Q(s_t, a)$
7:         Execute action $a_t$ in simulator and observe reward $r_t$ and state $s_{t+1}$
8:         $Q(s_t, a_t) = Q(s_t, a_t) + \alpha[r_t + \gamma \max_a Q(s_{t+1}, a) - Q(s_t, a_t)]$
9:         **if** $s_{t+1}$ is terminal **then**
10:            end episode
11:         **end if**
12:         $s_t = s_{t+1}$
13:     **end for**
14: **end for**

**Algorithm 2** Tabular Q-learning with UCB action selection

1: Initialize $Q(s, a)$, for all $s \in \mathcal{S}, a \in \mathcal{A}(s)$, arbitrarily
2: Initialize $N(s, a) = 0$, for all $s \in \mathcal{S}, a \in \mathcal{A}(s)$
3: **for** episode = 1, V **do**:
4:     $s_1$ = initial state
5:     **for** step = 1, T **do**:
6:         $a_t = \arg\max_a [Q(s_t, a) + c\sqrt{\frac{\ln N(s_t)}{N(s_t, a)}}]$
7:         Execute action $a_t$ in simulator and observe reward $r_t$ and state $s_{t+1}$
8:         $Q(s_t, a_t) = Q(s_t, a_t) + \alpha[r_t + \gamma \max_a Q(s_{t+1}, a) - Q(s_t, a_t)]$
9:         $N(s_t, a_t) = N(s_t, a_t) + 1$
10:         **if** $s_{t+1}$ is terminal **then**
11:            end episode
12:         **end if**
13:         $s_t = s_{t+1}$
14:     **end for**
15: **end for**

We implemented the DQL using experience replay and a seperate target neural network, as first presented by Mnih et al. [27]. The full algorithm is presented in Algorithm 3. This algorithm was chosen compared with the original DQL approach, which used only a single neural network, as it had improved performance in terms of learning rate and stability [27], [46]. We use a fully-connected single layer neural network for both the main and target neural networks, which takes a state vector as input and outputs the predicted value for each action (fig 4.3.1). The state vector is an ordered array of information for each machine on the network, while the output is an ordered array of values for each scan and exploit for each machine on the network. We utilize ε-greedy action selection for the DQL algorithm so that for a given state, the next action is chosen by selecting the action with the highest predicted value with probability 1 - ε, and a random action uniformly at random with ε probability.

---

**Algorithm 3** Deep Q-learning with experience replay
---
1: Initialize replay memory $\mathcal{D}$ to capacity $\mathcal{N}$
2: Initialize action-value function $Q$, with random weights $\theta$
3: Initialize target action-value function $\hat{Q}$, with weights $\hat{\theta} = \theta$
4: **for** episode $= 1$, V **do:**
5:    $s_1 =$ initial state
6:    **for** step $= 1$, T **do:**
7:       With probability $\epsilon$ select a random action $a_t$
8:       otherwise select $a_t = \arg\max_a Q(s_t, a; \theta)$
9:       Execute action $a_t$ in simulator and observe reward $r_t$ and state $s_{t+1}$
10:      Store transition $(s_t, a_t, r_t, s_{t+1})$ in $\mathcal{D}$
11:      Sample random minibatch of transitions $(s_j, a_j, r_j, s_{j+1})$ from $\mathcal{D}$
12:      Set $y_j = \begin{cases} r_j, & \text{if } s_j \text{ is terminal} \\ r_j + \gamma \max_{a'} \hat{Q}(s_{j+1}, a'; \hat{\theta}), & \text{otherwise} \end{cases}$
13:      Perform a gradient descent step on $(y_j - Q(s_j, a_j; \theta))^2$ with respect to weights $\theta$
14:      Every C steps reset $\hat{Q} = Q$
15:      **if** $s_{t+1}$ is terminal **then**
16:        end episode
17:      **end if**
18:      $s_t = s_{t+1}$
19:    **end for**
20: **end for**

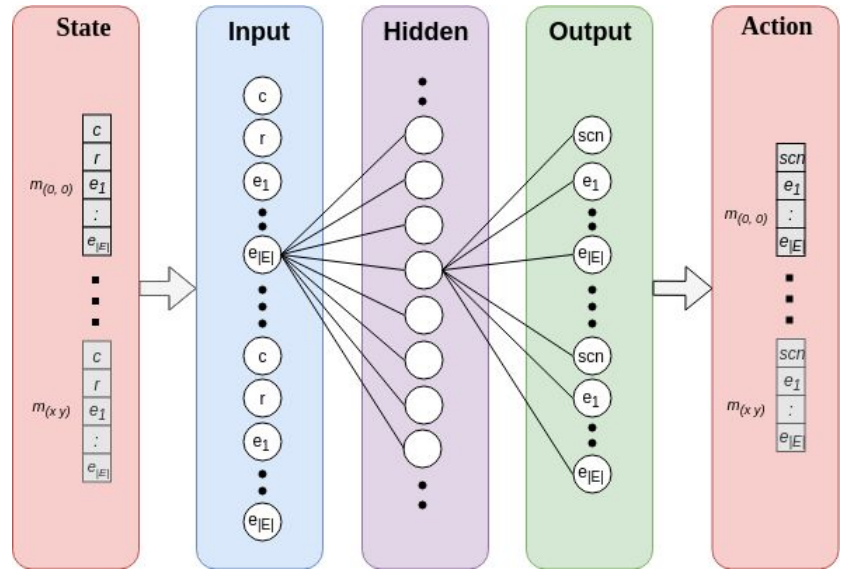

**Figure 4.3.1 | Schematic illustration of the Deep Q-network.** Input layer is a state represented by a 1D vector of information for each machine in network. The single hidden layer is fully connected to both input and output layers. The output layer is the action value for each scan and exploit for all machines in network.

## 4.4 Experiments

This study aimed to investigate the application of RL to Automated Penetration Testing. The main questions we wanted to answer were:

1) Can RL be used to find an attack path through a network when one exists?
2) How optimal is the attack path generated by the RL agent, as defined by the reward function?
3) How does RL scale with increased network size and number of exploits?
4) How general is the RL approach? Does it work on different network configurations?
5) How do the different RL approaches compare?

     For (1) an attack path, means the agent is able to learn a policy that is capable of reliably exploiting the sensitive machines on the network. For the case of our experiments (2) will be optimizing over the action cost. As discussed in chapter 3, action cost can be used to represent any desired property of an exploit, e.g. time required, monetary cost or chance of detection. For (3) network size means the number of machines, |M|, on the network. It is worth noting that both the state and action space grow with |M| and the number of exploitable services, |E|, so

increasing |M| or |E| increases the problem size for the RL agent. However, they both don't affect the dynamics of the system the same so it was important we investigated how performance was affected by both increasing |M| and |E|.

We investigated the performance of each RL algorithm by measuring performance on a range of scenarios. The general experimental procedure was to choose a scenario (i.e. network size, topology, number of services, etc..), train the agent for a set period of time and then evaluate the final policy generated by the agent. The following sections describe each of these steps in detail.

## Experiment scenarios

We tested the different RL algorithms on a number of different network scenarios using the NAS we developed. In particular we used three different computer network designs: i) the standard network described in chapter 3 ii) single site network and iii) multi-site wide-area network. Unfortunately, we were unable to find specific case study network designs to use for our scenarios apart from the standard network design which was based on practical commercial experience of the Sarraute et al. [12]. We therefore, designed the other two scenarios to represent a simple single location network and a multi-location network of the same size.

The standard network design is shown in figure 4.4.1 and described in detail in chapter 3. This architecture has been used in previous studies testing automated pentesting agents in simulated environments [11], [12], [38]. This design acts as a real-world scenario as well as a means of comparison. Additionally, the NAS supports generation of random scenarios using this design based on the number of machines and number of exploits. We use this feature when investigating the scaling properties of the different RL algorithms.

The single site and multi-site network scenarios are shown in figure 4.4.2 and 4.4.3. All three network scenarios contain 16 machines, five exploitable services and two sensitive machines. This was chosen so we could investigate the effect of different network architectures on the RL algorithm performance. Similarly, all scenarios require a minimum of three machines to be exploited in order to gain access to all the sensitive documents and complete the scenario.

The three fixed scenarios described are used to test questions (1), (2), (4) and (5) outlined at the start of this section. In order to investigate question (3) as well as (1) and (5), we tested the RL algorithms against the standard generated network scenarios while varying either the number of machines in the network and using a constant number of exploitable services or varying the number of exploitable services and keeping the number of machines constant .

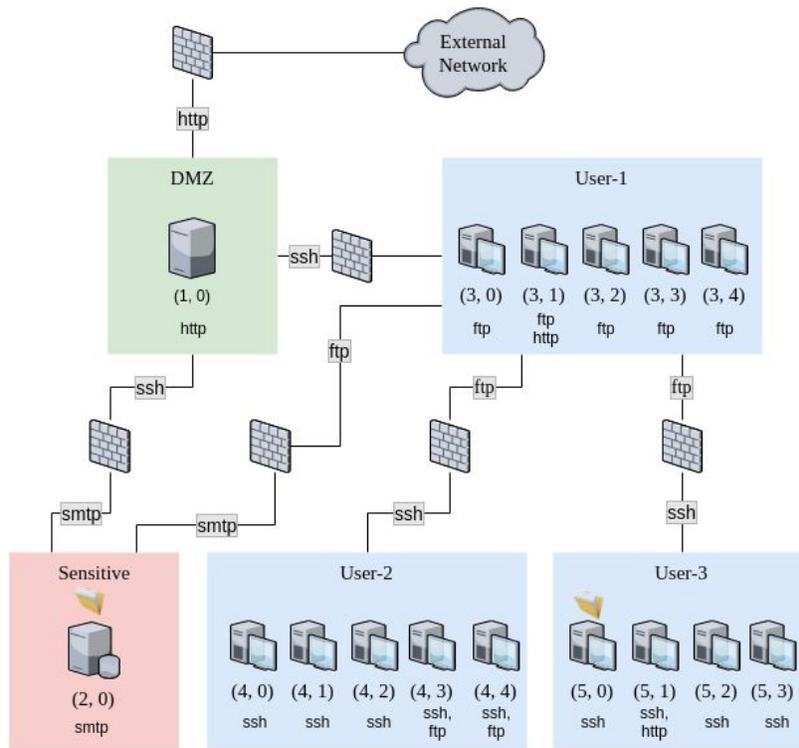

**Figure 4.4.1** | **The standard network scenario.** Labels beneath each machine indicate, the exploitable services running on that machine. Labels along edge to firewall from subnet indicates exploitable service traffic allowed through firewall. Documents above a machine indicate a valuable machine.

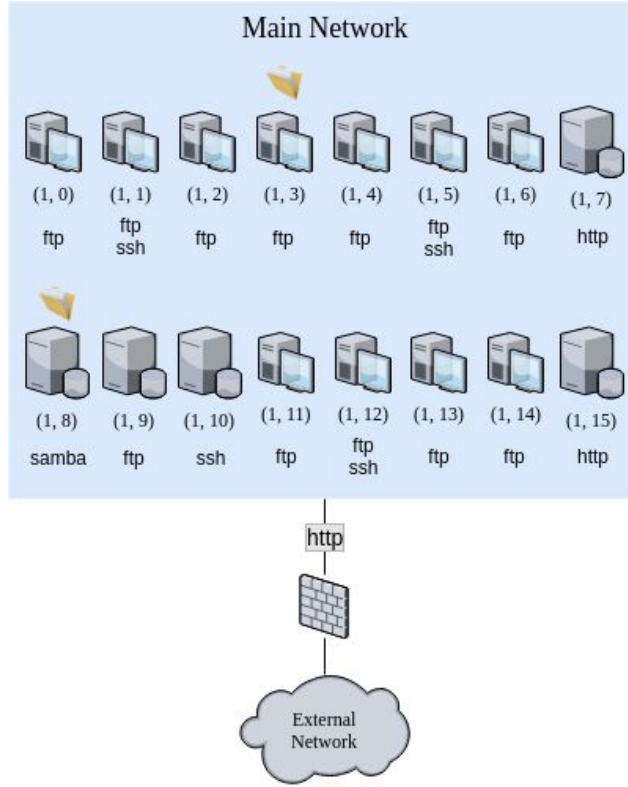

**Figure 4.4.2** | **The single site network scenario** (see figure 4.4.1 for diagram information).

For each scenario we tried to keep as many settings as consistent as possible, to better elucidate the effect of the variable of interest. Table 4.4.1, provides details of the different parameter values used. The max steps value chosen as this tended to give good performance across a range of scenario sizes during preliminary testing. Similarly for the values of sensitive machines and action costs.

**Table 4.4.1** | **Experiment scenario parameters and their values**

| Parameter | Value | Description |
| --- | --- | --- |
| Max steps | 500 | Maximum number of steps allowed per episode, before environment is reset to start state |
| Exploit cost | 1 | Cost of performing each exploit action |
| Scan cost | 1 | Cost of performing a scan action |
| Sensitive machine value | 10 | Reward received when sensitive machine is exploited |

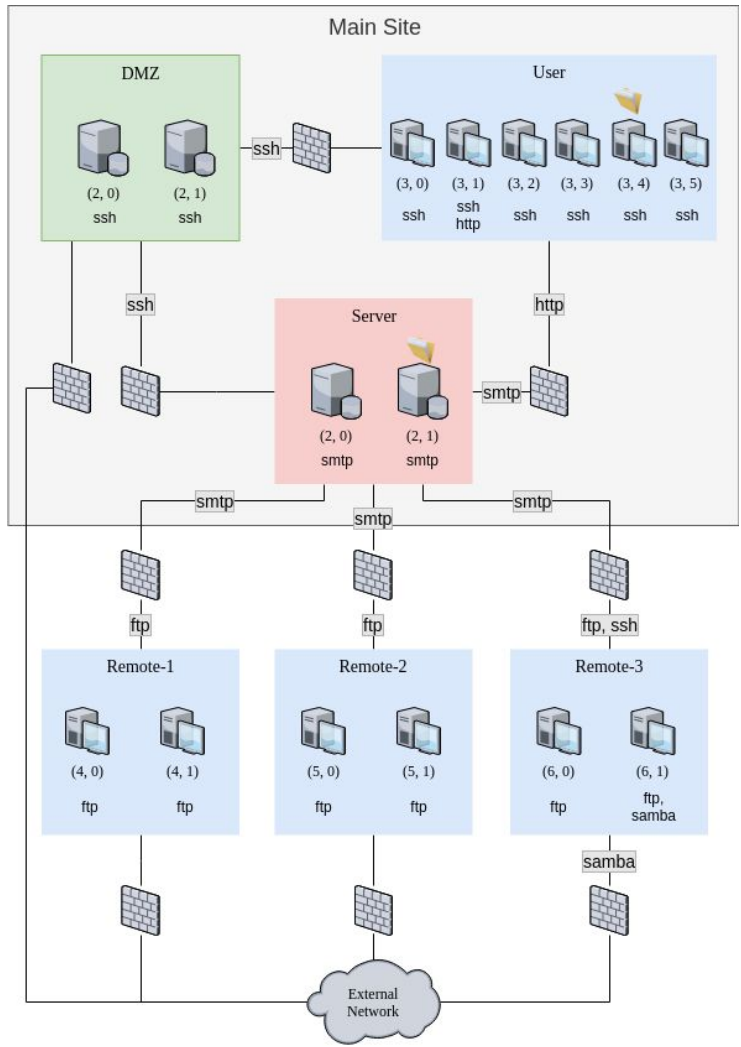

**Figure 4.4.3** | **The multi-site network scenario** (see figure 4.4.1 for diagram information).

Training details

    For each scenario we trained each agent for two minutes before evaluating the learned policy. We chose to train the agents for a set time period since the time per step of the tabular agents is significantly faster than the DQL algorithms and since, in practical terms, if applying RL in a commercial setting we would be most interested in how long it takes an agent to find an attack path in terms of time compared to training episodes. The time limit of two minutes was chosen for practical reasons, as it allowed a large enough variety of scenarios to be tested and solved while still being short enough to not make running of many experiments too time consuming.

Each algorithm required the selection of a number of hyperparameter values. The values chosen are shown in table 4.4.2. Hyperparameters selection was done using a mixture of informal search on different size standard generated networks, selecting default choices from the machine learning library used [37] and values seen in literature for DQL [27]. For tabular $\varepsilon$-greedy and DQL we also used $\varepsilon$-decay, which acts to reduce $\varepsilon$, and hence the probability of choosing a random action, over time from the initial value, $\varepsilon_{max}$, to the final value, $\varepsilon_{min}$ (1.0 and 0.05, respectively for both Tabular $\varepsilon$-greedy and DQL algorithms) (eq. 4.4).

$$\varepsilon_t = \varepsilon_{min} + (\varepsilon_{max} - \varepsilon_{min})\, e^{-\lambda t} \qquad (4.4)$$

**Table 4.4.2 | List of hyperparameters used for each algorithm and their values**

| Hyperparameter | Tabular $\varepsilon$-greedy | Tabular UCB | Deep Q-learning |
| --- | --- | --- | --- |
| Step size, $\alpha$ | 0.1 | 0.1 | - |
| Discount factor, $\gamma$ | 0.99 | 0.99 | 0.99 |
| Initial $\varepsilon$ value, $\varepsilon_{max}$ | 1.0 | - | 1.0 |
| Final $\varepsilon$ value, $\varepsilon_{min}$ | 0.05 | - | 0.05 |
| $\varepsilon$ decay rate, $\lambda$ | 0.0001 | - | 0.0001 |
| Confidence, $c$ | - | 0.5 | - |
| Minibatch size | - | - | 32 |
| Hidden layer size | - | - | 256 |
| Replay memory size | - | - | 10 000 |
| Target network update frequency | - | - | 1000 steps |
| RMSprop learning rate | - | - | 0.00025 |
| RMSprop gradient momentum | - | - | 0.9 |

### Evaluation procedure

Following training the agents performance was evaluated by running its trained policy against the network scenario in the NAS either 10 or 30 times depending on the experiment. The policy was tested using $\varepsilon$-greedy action selection with $\varepsilon = 0.05$, so to avoid the chance of the agent getting stuck in a state forever and to avoid overfitting of the policy to the scenario. For comparison, where applicable, we also used a random agent which selected actions uniformly at random in each state.

For the custom scenarios used (standard, single site and multi-site), we trained the agent once against each scenario and then ran 30 evaluations of the trained agent. For the experiments where we generated scenarios using the NAS, we did 10 separate runs for each scenario using a different random seed each run and evaluated each of the runs 10 times. We chose to generate multiple versions of the same scenario since performance varied significantly depending on the generated configurations, even for the same network topology. This is due to differences in the number of exploitable services available on any given machine and also the exploit probabilities, since both these factors were randomly generated.

### Experiment setup

All experiments were conducted on single core on a personal laptop computer running the Linux Ubuntu 18.04 operating system. The test machine was running an Intel Core i7 7th Gen CPU at 2.7 GHz and had 16GB of RAM. All RL algorithms were implemented in the Python 3 programming language (Python Software Foundation, www.python.org) and popular well-supported open source libraries. Specifically, the libraries used were numPy [33] for fast array based computation and Pandas [47] and Matplotlib [34] for handling and rendering of results. For the DQL algorithm we utilized the Keras Neural Network library [37], running on the TensorFlow library backend [36]. All Neural Network computation was done using a CPU only, with no GPU acceleration. Please see Appendix A for details on and access to source code for this project.

## 4.5 Results

### Custom Scenarios

For each of the different constructed network scenarios: i) standard ii) single site and iii) multi-site, we measured the episodic reward during training as well as final performance of the trained policy of each RL algorithm. Figure 4.5.1 shows the mean episodic reward versus the training episode. We averaged the reward over the last 100 episodes, in order to produce a smoother reward signal. All three algorithms converged on the approximate optimal solution within the training time limit, shown in the plots by the convergence of mean episodic reward to the theoretical max (red dashed line). The theoretical max is based on deterministic exploits and is the total value of the sensitive machines minus the total cost of the minimum number of exploits required to compromise them from the starting state. For the single-site and standard network scenarios, convergence occurred after a similar number of episodes for all three algorithms (~1000 episodes for single site and ~ 150 episodes for standard network). For the multi-site network scenario, the DQL algorithm converged significantly faster, after ~100 episodes compared with >1000 episodes for the two tabular algorithms.

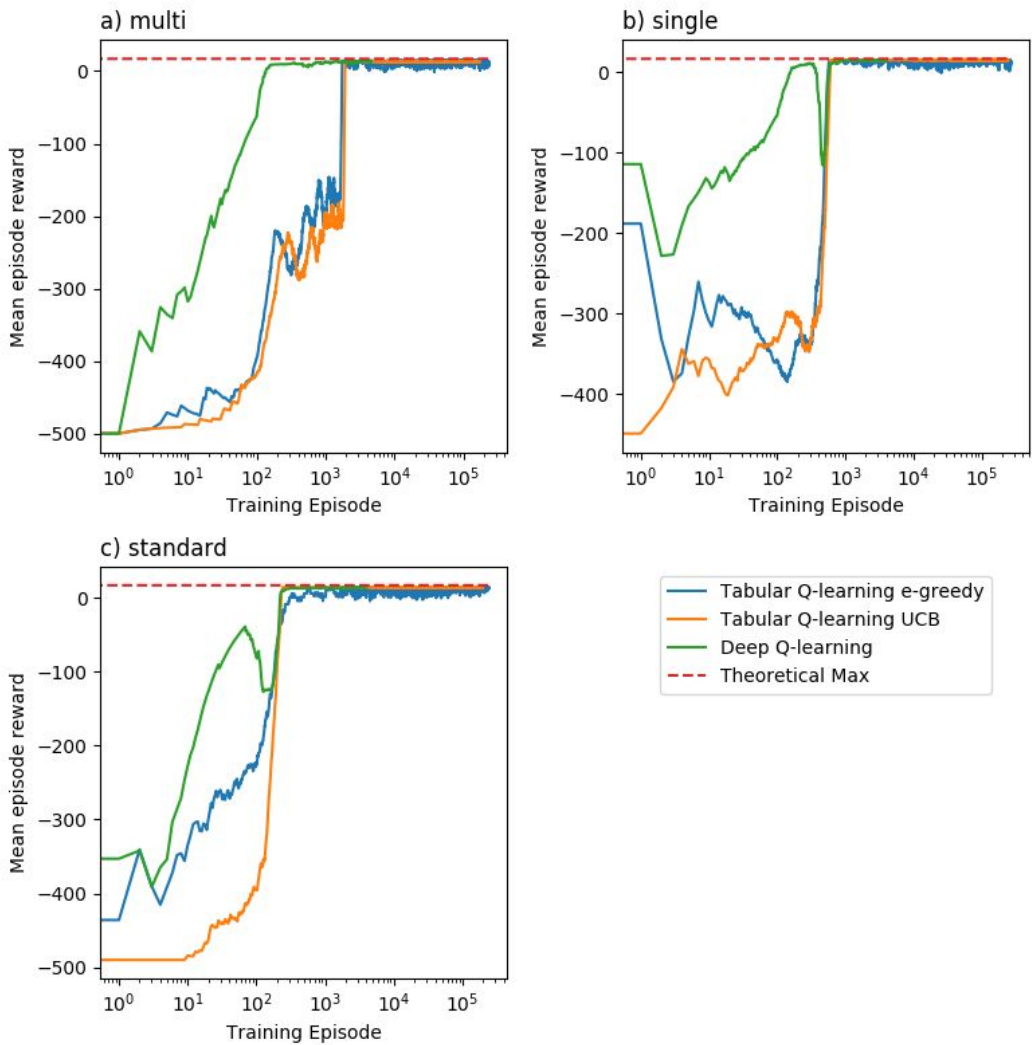

**Figure 4.5.1 | Mean episodic reward versus training episode** for the different RL algorithms and scenarios tested. The dotted red line represents the theoretical max reward obtainable for each scenario. The mean episodic reward is smoothed by averaging over last 100 episodes.

We also compared the mean episodic reward over time for the different RL algorithms (fig. 4.5.2). As opposed to reward versus episode where DQL tended to learn faster, the two tabular algorithms converged to the approximate optimal solution significantly faster than DQL in terms of time. For the single-site and standard network scenarios, the tabular methods converged in <10 seconds, while DQL took ~50 seconds for the standard network and ~75 seconds for the single-site network. For the multi-site network scenario, convergence time was more similar with all algorithms converging in <25 seconds, however DQL was still the slowest by ~5 seconds.

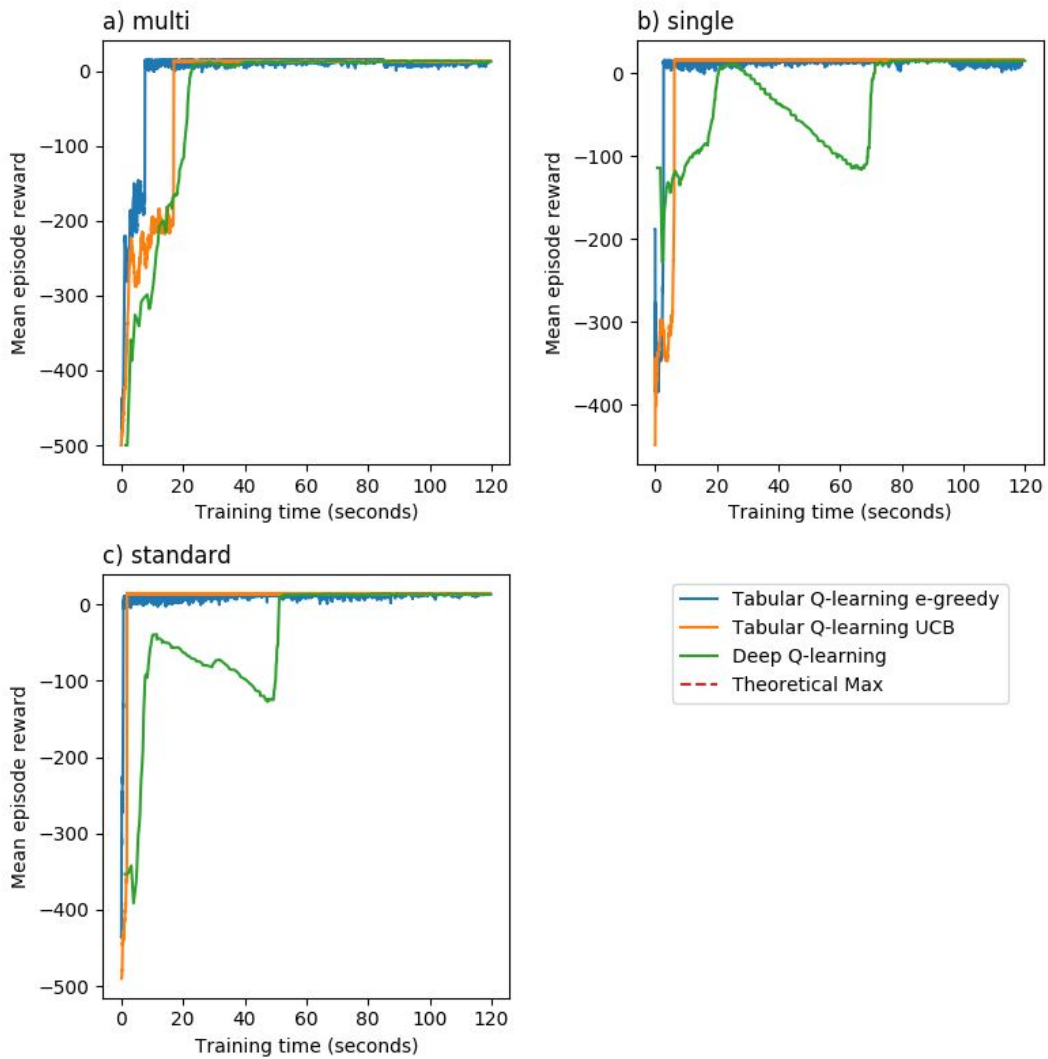

**Figure 4.5.2 | Mean episodic reward versus training time** (seconds) for the different RL agents and scenarios tested. The dotted red line represents the theoretical max reward obtainable for each scenario. The mean episodic reward is smoothed by averaging over last 100 episodes.

This difference in convergence time is likely related to the time each algorithm takes to complete an episode. To measure this we recorded the number of training episodes completed by each algorithm within the two minute training period (table 4.5.1). The tabular methods were significantly faster than the DQL algorithm, with Tabular $\varepsilon$-greedy performing 50 times more episodes than DQL in the worst case and Tabular UCB performing 37 times more episodes than DQL in the worst case. This difference is expected due to the additional computation required for the DQL Neural Network computations. The increased speed of the Tabular methods makes up for the slower learning rate per episode and still allows for them to have optimal performance in the tested scenarios.

**Table 4.5.1 | Number of training episodes for each Reinforcement Learning algorithm and scenario**, during two minute training period.

| Scenario | Tabular $\varepsilon$-greedy | Tabular UCB | Deep Q-Learning |
| --- | --- | --- | --- |
| Standard | 240 999 | 196 728 | 3959 |
| Single-site | 265 642 | 242 598 | 4016 |
| Multi-site | 234 042 | 171 942 | 4636 |

Following the two minute training period for each RL algorithm on each scenario we measured the performance of the final trained policies using $\varepsilon$-greedy action selection with $\varepsilon$ = 0.05. We recorded the proportion of the 30 evaluation runs which were solved by the agent, where a run was considered solved when the agent successfully exploited the sensitive machines on the network within the step limit (500 steps). We also recorded the maximum reward achieved by the agent for each scenario along with the variance in reward over the 30 evaluation runs. Performance of a random policy was also measured for comparison.

The results of the trained agent evaluations are shown in figure 4.5.3 and recorded in table 4.5.2. All three algorithms and the random policy were able to solve each scenario on at least some of the runs, with the three RL agents performing significantly better than random on the multi-site and standard network scenarios. For the single-site network scenario, the Tabular $\varepsilon$-greedy agent actually performed worse than the random agent (0.9 vs 0.97 respectively) while the Tabular UCB and DQL algorithms performed equally as good or better. The worse performance of Tabular $\varepsilon$-greedy is likely due to the $\varepsilon$-greedy action selection in the policy evaluation causing the agent to get stuck in a state not usually encountered along it's normal trajectory to the goal. Additionally, the performance of the random agent is very high for this scenario, due to its simplicity compared with the other scenarios. The only algorithm able to solve 100% of all evaluation runs was the DQL algorithm.

All algorithms were able to achieve an approximately optimal max reward for each scenario as shown by the plot of max reward in figure 4.5.3. The random agent was able to solve each scenario however it took significantly more actions each time, as shown by negative max reward achieved. The performance of the DQL algorithm was the most consistent across the different scenarios, likely attributable to the better generalization of it's policy to unseen states and states not along the optimal attack path.

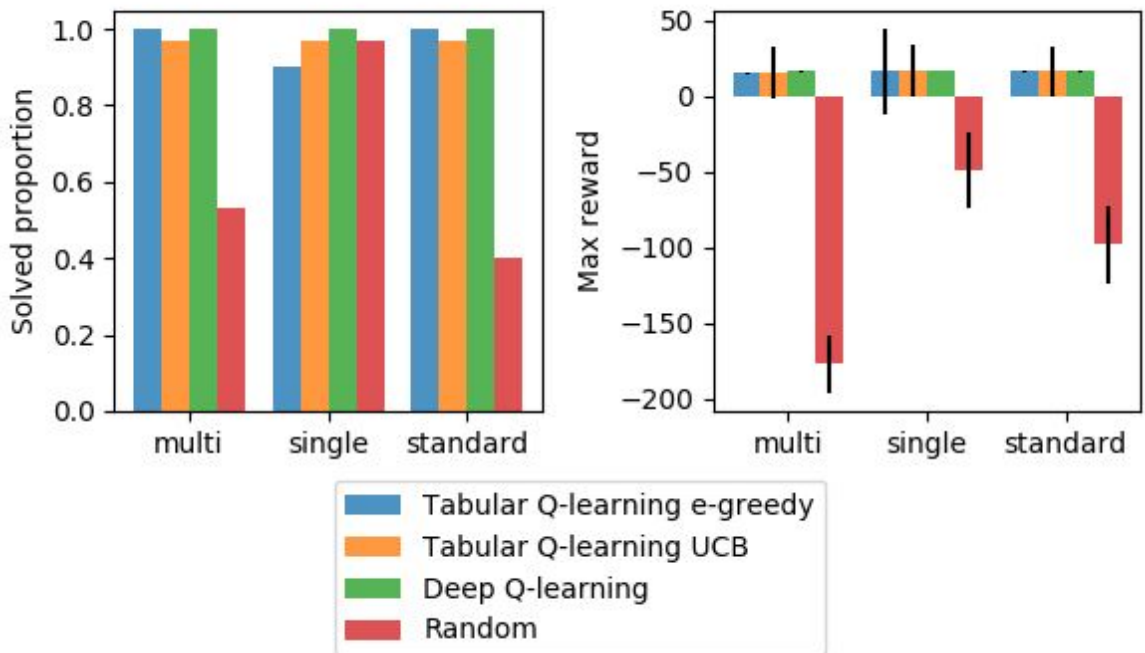

**Figure 4.5.3 | Evaluation performance of Reinforcement Learning algorithms for each scenario.** Solved proportion (left), shows proportion of 30 evaluation runs solved. The (right) plot shows the max reward achieved from the 30 evaluation runs, with error bars showing mean standard error of rewards.

**Table 4.5.2 | Solve proportion and max reward achieved for each scenario** and reinforcement learning algorithm following training. Solved proportion is top number while max reward (± mean standard error) is the lower number in each row.

| Scenario | Random | Tabular ε-greedy | Tabular UCB | Deep Q-Learning |
|---|---|---|---|---|
| Multi-site | 0.53<br>-177 (± 18.94) | 1.0<br>15 (± 0.38) | 0.97<br>15 (± 16.77) | 1.0<br>16 (± 0.30) |
| Single-site | 0.97<br>-49 (± 24.92) | 0.9<br>16 (± 28.64) | 0.97<br>17 (± 17.18) | 1.0<br>17 (± 0.24) |
| Standard | 0.4<br>-98 (± 25.45) | 1.0<br>16 (± 0.32) | 0.97<br>16 (± 16.82) | 1.0<br>16 (± 0.3) |

Scaling

We measured the scaling performance of each algorithm in terms of network size and number of exploits. We measured the effects of both using the standard generated network of the NAS. For each network scenario we tested 10 different random configurations and then measured performance on each configuration following two minutes training using 10 evaluation runs with ε-greedy action selection with $\varepsilon = 0.05$. We measured performance using two metrics. First was mean solved proportion, which was an average of the proportions of evaluation runs solved for each different configuration, where a network was solved when all sensitive machines were exploited within the step limit of 500. The second metric was mean reward, which was the average over runs of the reward achieved during evaluation runs. We also evaluated a random policy for a comparison.

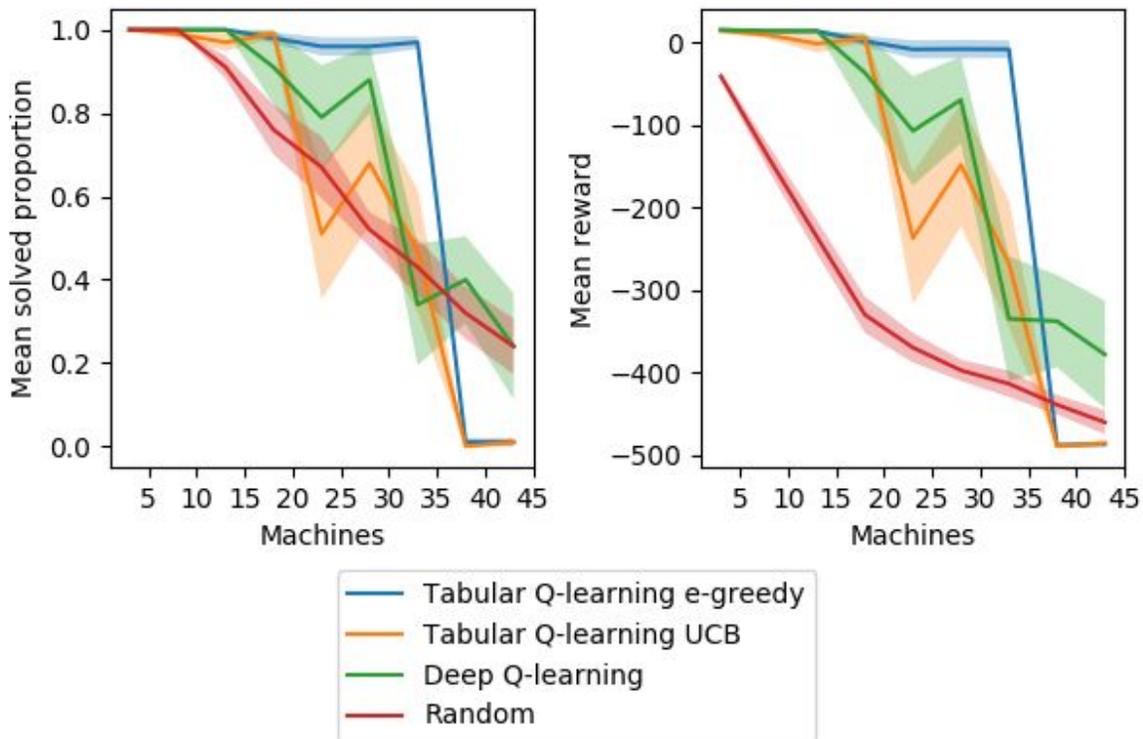

**Figure 4.5.4 | Reinforcement Learning algorithm performance versus number of machines** in auto-generated standard networks with number of exploitable services fixed at 5. For each network size tested, performance was averaged over 10 evaluation runs of trained agents using an ε-greedy policy with ε = 0.05 for 10 different runs for each scenario, where machine configurations change between runs. The shaded areas show mean standard error.

The effect of network size was tested by increasing the number of machines on the network from 3 to 43 machines in intervals of 5, while keeping the number of exploitable services fixed at 5. Figure 4.5.4 shows the results of the experiments. Performance was equal and better than a random policy for all three algorithms up to networks containing 18 machines. For the tested networks with more than 18 machines, performance of DQL and Tabular UCB algorithms declined rapidly, although both algorithms were still able solve more than 50% of scenarios for networks with 23 and 28 machines with mean reward performance still better than random. Performance for Tabular ε-greedy was consistently high for network up to and including 33 machines, after which performance rapidly dropped to worse than random for networks with 43 machines.

We measured the effect of increasing the number of exploits available to the agents by using a fixed size network of 18 machines and increasing the exploitable services available from 1 to 51 in intervals of 5. The results of the experiment is shown in figure 4.5.5. The effect of increased number of exploits differed for each algorithm. Performance was relatively unaffected for the Tabular ε-greedy, which maintained near optimal performance for all numbers of exploits tested. Similarly, Tabular UCB, had lower than optimal performance, but performance remained relatively consistent as the number of exploits increased. The most affected was DQL which had performance comparable with random for all values tested (we leave discussion of why this may have been the case to the next section).

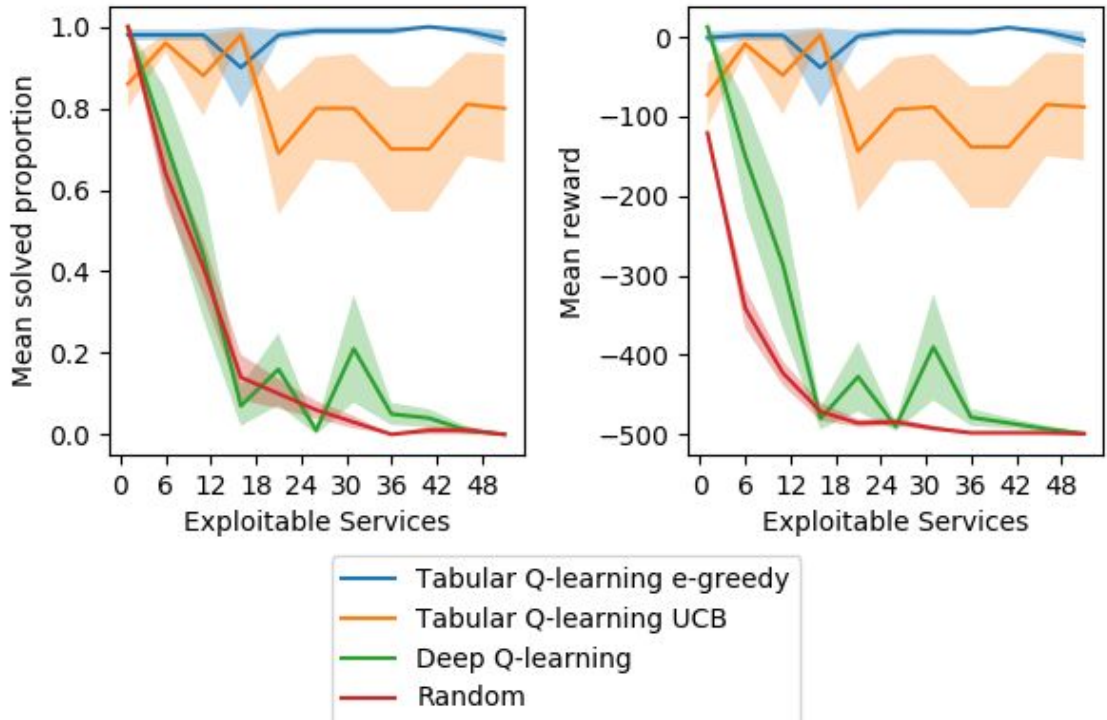

**Figure 4.5.5 | Reinforcement Learning algorithm performance versus number of exploitable services** in auto-generated standard networks with number of machines in network fixed at 18. For each network setting tested, performance was averaged over 10 evaluation runs of trained agents using an ε-greedy policy with ε = 0.05 for 10 different runs for each scenario, where machine configurations change between runs. The shaded areas show mean standard error.

## 4.6 Discussion

This aim of this study was to investigate the application of RL to pentesting. We did this by developing and testing a number of RL algorithms using the NAS we developed. In order to assess the applicability of RL we broke the problem down into five avenues of investigation. Specifically:

1) Can RL be used to find an attack path through a network when one exists?
2) How optimal is the attack path generated by the RL agent, as defined by the reward function?
3) How does RL scale with increased network size and number of exploits?
4) How general is the RL approach? Does it work on different network configurations?
5) How do the different RL approaches compare?

We aimed to answer questions (1), (2) and (4) by constructing some custom scenarios and then measuring the performance of the attack policies generated by trained RL agents in those scenarios. The scenarios are shown in figures 4.4.1-3, while the results of our experiments are shown in 4.5.1-3 and tables 4.5.1-2. For each scenario all RL algorithms tested were able to find an attack policy that lead to all sensitive machines on the target network being compromised, while also minimizing the number of actions taken, performing significantly better than a random policy in terms of max reward achieved. Based on these results, within the simulated environment RL is able to find a valid attack path (1) that is also optimal in the reward achieved and hence action cost (2) and is also able to be applied to different network configurations (4).

Although these results provide promising evidence for the use of RL in automated pentesting, it is worth noting some limitations of the experiments. In particular, the scenarios were were limited to networks containing 16 machines with five exploitable services and so the were is relatively small in size compared with large commercial networks of hundreds of machines that can be found in real-world settings. However, along these lines, the agents were only allowed two minutes of training with the slowest time to convergence being roughly 80 seconds for the DQL algorithm (fig. 4.5.2). It would therefore be expected that the RL agents could be applied to larger problem sizes given more training time. Another limitation is that we only tested three different network topologies, while the possible topologies encountered in the real-world are endless. One of the key advantages of RL is its generality, as demonstrated by its success in achieving human-level or better performance across a wide range of very distinct games using a single algorithm design [27], [46]. Based on this fact it would be expected that RL would be able to generalize to other network topologies, since for the RL algorithms the difference in underlying problem structure and complexity between different network topologies is relatively small, the key difference comes in the difference in size of the networks.

To investigate how RL performance differed with problem size we looked at the scalability of RL (3). The scalability of the different RL algorithms was measured both in terms of number of machines on the network (fig. 4.5.4) and also the number of exploits available to the agent (fig. 4.5.5). We found that for all the algorithms tested their performance dropped rapidly beyond a certain network size (18 machines for Tabular UCB and DQL, and 38 for Tabular ε-greedy). The drop in performance is expected since for the pentesting problem, how we have defined it, the size of the state space grows exponentially with the number of machines on the network (eq. 3.1) while the training time was kept constant at two minutes. For larger networks, increasing training time would likely lead to improved performance, at least up to a point. Performance of Tabular RL algorithms is known to suffer when the state space grows really large since they do not generalize action-values between similar states [14]. This can be seen in the steep drop in performance to below that of a random policy of the Tabular RL algorithms for networks with 43 machines (fig 4.5.4), while the DQL algorithm is still able to outperform a random policy for this size network. It would be expected using DQL or related function-approximation approaches to RL it should be possible to scale up these algorithms to deal with much larger networks given enough training time. As an example of the scalability of DQL approaches, the AlphaGo RL agent is able to find better than human level performance in a game with a state space of around $10^{174}$ states [15].

Another approach to handling the exponential increase in the state space size would be to decompose the problem into two seperate agents. One agent would work at the network topology level, deciding which machine to target next, while the second agent works at the individual machine level, deciding which exploits to use against a given machine. This approach has been shown to work well for model-based automated pentesting using POMDPs [12]. One promising approach for model-free automated pentesting would be the use of hierarchical RL, which is a technique that can be applied to MDP to decompose actions into different levels of abstractions, however this is still an area of active research in the RL domain and so it may be some time before it becomes a reliable and efficient method [48]. Between the possibility of scaling up RL via more efficient function-approximation methods or via problem decomposition, it is possible to imagine RL being applied to very large networks.

Apart from large network size, another of the challenges facing pentesting is the complexity and rapid evolution of the services present on a given system and the set of exploits available. Every year new exploits are found for software and so in order to be useful automated pentesting agents will need to be able to handle a large growing database of exploits. We investigated the performance of RL algorithms as the number of available exploits increased (fig. 4.5.5). We found that the performance of the tabular RL algorithms were relatively unaffected by increased number of exploits, with the Tabular $\varepsilon$-greedy algorithm maintaining near optimal performance and solving almost 100% of every scenario configuration tested. On the other hand the performance of DQL was not much better than a random policy.

The reasoning for this reduced performance in DQL with increasing number of exploits we believe is due to two reasons. Firstly, increasing the number of actions increases the size of the neural network used by the DQL algorithm, which slows down the number of operations per second and hence the number of steps the algorithm can perform within the two minute training period. This is coupled with the fact the state space grows as the number of exploits increases, so slower exploration with a larger space to explore, means the agent cannot explore sufficiently to find the optimal attack plan within the time limit. The second key reason is that unlike states which the neural network is able to generalize well, since as the state space grows so does the number of very similar states as many machines on the network can be ignored as they are not along the optimal path. However, as the number of exploits increases, so do the number of state possibilities along the optimal path since the branching factor for each machine grows with the number of exploits. Decomposing the automated penetration testing problem into a higher level topology agent and a machine exploiting agent would help alleviate this problem, since then adding a new exploit would only mean a single increase in state space size for the machine level agent, rather than an exponential increase as is the case with the current design.

Based on these scaling experiments, it would appear that the approach used in this study would not scale well to large networks and number of exploits. The Tabular RL algorithms are able to scale well with the number of actions, while not being able to scale to networks with many machines. Conversely, it may be possible to scale the DQL algorithm to larger networks given more training time and using more advanced algorithms, however this approach does not scale well with increased number of actions. In terms of action selection strategy, $\varepsilon$-greedy action selection learned faster, produced better policies for the majority of scenarios tested and scaled significantly better than UCB for the tabular algorithms, so is a clear choice for future research and application.

## 4.7 Conclusion

This study was a preliminary investigation into the application of RL to pentesting. We demonstrated that in a simulated environment it is possible to use RL to generate a policy that is capable of successfully and optimally exploiting a target network. Taking this a step further, it would be possible to use the policy information to determine which components of a network need patching and then repeat the process in order to improve the security of a network. The key advantage of using RL over other AI planning techniques, is that it requires no prior knowledge of the transition model of the exploits and network and so is very general in its application, with the same algorithm able to be applied to different network topologies and configurations with varying number of exploits available.

Although RL will be a powerful tool for many domains as technology improves, for automated pentesting there are still a number of hurdles to overcome. In particular, scaling the algorithms with the number of machines on the network and also with the number of exploits available to the agents. Further work into developing more efficient algorithms or decomposing the pentesting problem will help greatly to solve these issues.

# Chapter 5

# Conclusions

## 5.1 Summary and conclusions

In this project we wanted to investigate solutions for the challenging problem of cybersecurity, in particular the growing shortage of skilled professionals. As with many industries, automation is an obvious choice for a solution when tackling a skills shortage. We aimed to worked on applying the AI approach of RL to automating network pentesting. This involved two stages, the first of which was to design and build a NAS that could be used to design and test automated pentesting agents in a fast easy to use environment. The second stage was to utilize the NAS to study the application of RL algorithms to automated pentesting.

The main finding of this work, was that within a simulated environment it is possible for RL algorithms to find optimal attack paths through a target computer network. We were able to successfully apply RL to a range of network topologies and configurations with the only prior knowledge being the topology of the network and the set of available scans and exploits. This offers a key advantage over current approaches to automated pentesting which rely on having an accurate model of the exploit outcomes, which is hard to produce in the rapidly evolving cyber security landscape. However, we also found that the RL algorithms we implemented are limited in their applicability to relatively small networks and number of exploits. Additionally, RL relies heavily on having a high-fidelity simulator for training agents before they are capable of being used in a real-world scenario.

## 5.2 Possible future work

This study aimed to be a preliminary study into using RL for automated pentesting and as such has lead to many more interesting problems to solve before this technology may be practically used in a commercial environment. Firstly, we need to develop more scalable RL algorithms that are able to scale to the size of modern large networks while also handling hundreds if not thousands of possible exploits. The next step after that will be to apply these algorithms in more realistic environments such as VM networks using information from real organizational networks in order to determine how they can be applied in real-world settings.

# Appendix A

# Program listing

All the source code and results for this project can be found at the github repo:
https://github.com/Jjschwartz/NetworkAttackSimulator.

The code for this project is all written in Python 3 and is split into a number of separate modules. The main modules are the Environment, the Agents and the Experiments. There is also a module for generating a POMDP file, however this was not used for the work covered in this thesis and so is ignored for the sake of clarity.

## A.1 Environment

This module contains all the code for the Network Attack Simulator. The program architecture is shown in figure 3.3.1 in Chapter 3 of this thesis. Table A1 provides a list of each file in the module and a description of its role in the Network Attack Simulator.

**Table A1 | Files and description in the program Environment module.**

| File | Description |
| --- | --- |
| action.py | Contains the Action class which defines an action in the NAS |
| environment.py | The main class for the NAS. It controls the network attack model and provides the main function with which to interact with the NAS |
| generator.py | Contains all the functionality for automatically generating a network configuration given the number of machines and services to run. |
| loader.py | Contains all functionality for loading a custom network configuration from a .yaml file |
| machine.py | Contains the Machine class which defines a machine in the network model |
| network.py | The network model class, which defines and controls the behaviour and configuration of the network in the NAS |
| render.py | Contains functionality for rendering the environment and episodes |
| state.py | Contains the State class which defines a state in the NAS |

## A.2 Agents

This module contains all the different Q-learning agents that were implemented for this project. Table A2 provides a description of all files in this module.

**Table A2 | Files and description in the program Agent module.**

| File | Description |
| --- | --- |
| agent.py | Contains the abstract Agent class which all RL agents inherit from |
| dqn.py | Contains the Deep Q-Learning agent class |
| q_learning.py | Contains the Tabular Q-learning agents (both $\varepsilon$-greedy and UCB) |
| random.py | Contains the Random agent |
| tabular.py | Contains the abstract TabularAgent class which the Tabular Q-learning agents inherit from |
| tuner.py | Contains some functionality for performing hyperparameter search on the different agents |

# A.3 Experiments

This module contains all the different scripts used to run the experiments on the RL agents and NAS. Table A3 provides a description of all files in this module.

**Table A3 | Files and description in the program Experiment module.**

| File | Description |
| --- | --- |
| agent_perf_eval_plotter.py | Contains functionality for summarizing and displaying the results generating by evaluating trained agents using agent_perf_exp.py |
| agent_perf_exp.py | Runs performance and scaling experiments on RL agents. |
| agent_perf_plotter.py | Contains functionality for plotting episodic performance of RL agents |
| agent_scaling_plotter.py | Contains functionality for plotting the scaling performance of RL agents |
| env_perf_exp.py | Contains some functionality for testing the running performance of the NAS |
| env_perf_plotter.py | Functionality for plotting the results of experiments run on NAS |
| experiment_util.py | Contains functionality for loading scenarios and agents common across different experiments |
| results/ | Contains the main result files from the study |

# Bibliography

[1] T. J. Holt, O. Smirnova, and Y. T. Chua, "Exploring and Estimating the Revenues and Profits of Participants in Stolen Data Markets," *Deviant Behav.*, vol. 37, no. 4, pp. 353–367, Apr. 2016.
[2] Symantec Corporation, "Internet Security Threat Report," Volume 22, Apr. 2017.
[3] Australian Cyber Security Centre, "ACSC Threat Report," Australian Government, 2017.
[4] R. von Solms and J. van Niekerk, "From information security to cyber security," *Comput. Secur.*, vol. 38, pp. 97–102, Oct. 2013.
[5] D. of Industry, "$50 million investment into cyber security research and industry solutions," *Ministers for the Department of Industry, Innovation and Science*, 22-Sep-2017. [Online]. Available: http://minister.industry.gov.au/ministers/craiglaundy/media-releases/50-million-investment-cyber-security-research-and-industry. [Accessed: 13-Mar-2018].
[6] B. Arkin, S. Stender, and G. McGraw, "Software penetration testing," *IEEE Secur. Priv.*, vol. 3, no. 1, pp. 84–87, Jan.-Feb 2005.
[7] A. IVan, "Why Attacking Systems Is a Good Idea," *IEEE Secur. Priv.*, 2004.
[8] Cisco, "Mitigating the Cyber Skills Shortage," CISCO, 2015.
[9] C. Sarraute, *Automated Attack Planning*. 2012.
[10] C. Phillips and L. P. Swiler, "A graph-based system for network-vulnerability analysis," in *Proceedings of the 1998 workshop on New security paradigms*, 1998, pp. 71–79.
[11] C. Sarraute, O. Buffet, and J. Hoffmann, "Penetration Testing == POMDP Solving?," in *SecArt*, 2011.
[12] C. Sarraute, O. Buffet, and J. Hoffmann, "POMDPs Make Better Hackers: Accounting for Uncertainty in Penetration Testing," *AAAI*, 2012.
[13] J. Hoffmann, "Simulated penetration testing: From 'dijkstra' to 'turing test++,'" presented at the Proceedings International Conference on Automated Planning and Scheduling, ICAPS, 2015, vol. 2015-January, pp. 364–372.
[14] R. S. Sutton and A. G. Barto, "Reinforcement learning: An introduction," 2011.
[15] D. Silver *et al.*, "Mastering the game of Go without human knowledge," *Nature*, vol. 550, no. 7676, pp. 354–359, Oct. 2017.
[16] J. Kober, J. A. Bagnell, and J. Peters, "Reinforcement learning in robotics: A survey," *Int. J. Rob. Res.*, vol. 32, no. 11, pp. 1238–1274, Sep. 2013.
[17] R. R. Linde, "Operating system penetration," in *Proceedings of the May 19-22, 1975, national computer conference and exposition*, 1975, pp. 361–368.
[18] F. Holik, J. Horalek, O. Marik, S. Neradova, and S. Zitta, "Effective penetration testing with Metasploit framework and methodologies," presented at the CINTI 2014 - 15th IEEE International Symposium on Computational Intelligence and Informatics, Proceedings, 2014, pp. 237–242.
[19] G. Lyon, *Nmap Network Scanning: Official Nmap Project Guide to Network Discovery and Security Scanning*. Insecure.Com, LLC, 2008.
[20] P. Ammann, D. Wijesekera, and S. Kaushik, "Scalable, graph-based network vulnerability analysis," in *Proceedings of the 9th ACM conference on Computer and communications security*, 2002, pp. 217–224.
[21] M. S. Boddy, J. Gohde, T. Haigh, and S. A. Harp, "Course of Action Generation for Cyber Security Using Classical Planning," in *ICAPS*, 2005, pp. 12–21.
[22] K. Durkota and V. Lisý, "Computing Optimal Policies for Attack Graphs with Action Failures and

Costs," in *STAIRS*, 2014, pp. 101–110.

[23] R. Bellman, "A Markovian Decision Process," *Journal of Mathematics and Mechanics*, vol. 6, no. 5, pp. 679–684, 1957.

[24] G. E. Monahan, "State of the art—a survey of partially observable Markov decision processes: theory, models, and algorithms," *Manage. Sci.*, vol. 28, no. 1, pp. 1–16, 1982.

[25] A. R. Cassandra, *Exact and approximate algorithms for partially observable Markov decision processes*. Brown University, 1998.

[26] M. Mundhenk, J. Goldsmith, C. Lusena, and E. Allender, "Complexity of finite-horizon Markov decision process problems," *J. ACM*, vol. 47, no. 4, pp. 681–720, 2000.

[27] V. Mnih *et al.*, "Human-level control through deep reinforcement learning," *Nature*, vol. 518, p. 529, Feb. 2015.

[28] M. G. Bellemare, Y. Naddaf, J. Veness, and M. Bowling, "The Arcade Learning Environment: An Evaluation Platform for General Agents," *1*, vol. 47, pp. 253–279, Jun. 2013.

[29] J. Deng, W. Dong, R. Socher, L. Li, K. Li, and L. Fei-Fei, "ImageNet: A large-scale hierarchical image database," in *2009 IEEE Conference on Computer Vision and Pattern Recognition*, 2009, pp. 248–255.

[30] G. F. Riley and T. R. Henderson, "The ns-3 Network Simulator," in *Modeling and Tools for Network Simulation*, K. Wehrle, M. Güneş, and J. Gross, Eds. Berlin, Heidelberg: Springer Berlin Heidelberg, 2010, pp. 15–34.

[31] B. Lantz, B. Heller, and N. McKeown, "A network in a laptop: rapid prototyping for software-defined networks," in *Proceedings of the 9th ACM SIGCOMM Workshop on Hot Topics in Networks*, 2010, p. 19.

[32] A. Futoransky, F. Miranda, J. Orlicki, and C. Sarraute, *Simulating Cyber-Attacks for Fun and Profit*. 2010.

[33] T. E. Oliphant, *Guide to NumPy*. USA: Trelgol Publishing, 2006.

[34] J. D. Hunter, "Matplotlib: A 2D Graphics Environment," *Comput. Sci. Eng.*, vol. 9, no. 3, pp. 90–95, 2007.

[35] A. A. Hagberg, D. A. Schult, and P. J. Swart, "Exploring Network Structure, Dynamics, and Function using NetworkX," in *Proceedings of the 7th Python in Science Conference*, 2008, pp. 11–15.

[36] Martín Abadi *et al.*, "TensorFlow: Large-Scale Machine Learning on Heterogeneous Systems." 2015.

[37] F. Chollet and Others, "Keras," 2015. [Online]. Available: https://keras.io.

[38] M. Backes, J. Hoffmann, R. Künnemann, P. Speicher, and M. Steinmetz, "Simulated Penetration Testing and Mitigation Analysis," *arXiv:1705.05088 [cs]*, May 2017.

[39] S. Donnelly, " 'Soft Target: The Top 10 Vulnerabilities Used by Cybercriminals," Recorded Future, 2018.

[40] P. Mell, K. Scarfone, and S. Romanosky, "Common Vulnerability Scoring System," *IEEE Secur. Priv.*, vol. 4, no. 6, pp. 85–89, 2006.

[41] First, "Common Vulnerability Scoring System SIG," *First*. [Online]. Available: https://www.first.org/cvss/. [Accessed: 31-Oct-2018].

[42] Oracle, *Oracle VM VirtualBox*. 2017.

[43] Rapid, *Metasploitable 2 Virtual Machine*. 2015.

[44] C. J. Watkins and P. Dayan, "Q-learning," *Mach. Learn.*, vol. 8, no. 3–4, pp. 279–292, 1992.

[45] P. Auer, N. Cesa-Bianchi, and P. Fischer, "Finite-time Analysis of the Multiarmed Bandit Problem," *Mach. Learn.*, vol. 47, no. 2, pp. 235–256, May 2002.

[46] V. Mnih *et al.*, "Playing Atari with Deep Reinforcement Learning," *arXiv [cs.LG]*, 19-Dec-2013.

[47] W. McKinney, "Data Structures for Statistical Computing in Python," in *Proceedings of the 9th*

*Python in Science Conference*, 2010, pp. 51–56.

[48] R. S. Sutton, D. Precup, and S. Singh, "Between MDPs and semi-MDPs: A framework for temporal abstraction in reinforcement learning," *Artif. Intell.*, vol. 112, no. 1, pp. 181–211, Aug. 1999.

[47] R. S. Sutton, D. Precup, and S. Singh, "Between MDPs and semi-MDPs: A framework for temporal abstraction in reinforcement learning," *Artif. Intell.*, vol. 112, no. 1, pp. 181–211, Aug. 1999.